\documentclass[a4paper,titlepage,11pt]{article}
\pdfoutput=1
\usepackage[utf8]{inputenc}
\usepackage{geometry}
\usepackage[english]{babel}
\usepackage{amsmath, amsfonts, graphicx}
\usepackage{mathtools}
\usepackage{graphicx}
\usepackage{float}
\usepackage{pstricks}
\usepackage{hyperref}
\usepackage[labelfont=bf,font={small}]{caption}
\usepackage{pgfplots}
\usepackage{tikz}
\usetikzlibrary{positioning, arrows}
\usetikzlibrary{external}
\usepackage{footmisc}
\usepackage{multirow}
\usepackage{url}
\usepackage[breakable, theorems, skins]{tcolorbox}
\usepackage{enumerate}
\usepackage{physics}
\usepackage{bm}
\usepackage{epstopdf}

\numberwithin{equation}{section}

\newcommand{\normal}[1]{\hspace{0.2cm}: \hspace{-0.1cm} #1 \hspace{-0.1cm}:\hspace{0.2cm}}
\newcommand{\normaltwo}[1]{: \hspace{-0.1cm} #1 \hspace{-0.1cm}:}

\begin{document}
\begin{titlepage}

\setcounter{page}{1} \baselineskip=15.5pt \thispagestyle{empty}

%\begin{flushright}
%hep-th/13mmnnn\\
%\end{flushright}
\vfil

${}$
\vspace{1cm}

\begin{center}

\def\thefootnote{\fnsymbol{footnote}}
\begin{changemargin}{0.05cm}{0.05cm} 
\begin{center}
{\Large \bf Towards Black Hole Evaporation in Jackiw-Teitelboim Gravity}
\end{center} 
\end{changemargin}

~\\[1cm]
{Thomas G. Mertens\footnote{\href{mailto:thomas.mertens@ugent.be}{\protect\path{thomas.mertens@ugent.be}}}}
\\[0.3cm]

{\normalsize { \sl Department of Physics and Astronomy
\\[1.0mm]
Ghent University, Krijgslaan, 281-S9, 9000 Gent, Belgium}}\\[3mm]

\end{center}

%\vspace{2cm}
%\vspace{1cm}

%\hrule
 \vspace{0.2cm}
\begin{changemargin}{01cm}{1cm} 
{\small  \noindent 
\begin{center} 
\textbf{Abstract}
\end{center} 
Using a definition of the bulk frame within 2d Jackiw-Teitelboim gravity, we go into the bulk from the Schwarzian boundary. Including the path integral over the Schwarzian degrees of freedom, we discuss the quantum gravitational Unruh effect and the Planckian black-body spectrum of the thermal atmosphere. We analyze matter entanglement entropy and how the entangling surface should be defined in quantum gravity. Finally, we reanalyze a semi-classical model for black hole evaporation studied in \cite{ads2} and compute the entanglement between early and late Hawking radiation, illustrating information loss in the semi-classical framework. 
}
\end{changemargin}
 \vspace{0.3cm}
%\hrule
\vfil
\begin{flushleft}
\today
%March 20, 2013
\end{flushleft}

\end{titlepage}
\newpage
\tableofcontents
%\newpage

\setcounter{footnote}{0}

\section{Introduction}
Understanding black hole evaporation and information loss is one of the big unsolved problems in quantum gravity \cite{Hawking:1976ra}. \\
Ever since the advent of AdS/CFT it has been understood that the evaporation process is unitary since it can be described by a unitary QM theory on the boundary, and hence information cannot be lost. However, this state of affairs is unsatisfactory, as we still have no precise understanding of the bulk mechanism that releases the quantum information during evaporation. \\
One of the hurdles we are faced with here is the difference in language between the two set-ups. Hawking's computation is done in the context of quantum field theory in curved spacetimes, neglecting the matter-gravitational interactions and hence backreaction in the process. Boundary holography on the other hand faces the problem of bulk reconstruction: how does one construct local bulk operators and make contact with local bulk dynamics, as would be constructed by e.g. an infalling observer.
\\~\\
In order to address these problems, it is useful to have interesting toy models where aspects of these computations can be carried out exactly. An interesting model studied primarily in the '90s is the 2d CGHS model \cite{Callan:1992rs,Giddings:1992ff,Schoutens:1993hu,Almheiri:2013wka}. Explicit computations can be performed probing semi-classical Hawking evaporation, but it is asymptotically flat and hence more difficult to embed within a tractable unitary framework. Another model that has attracted a considerable amount of attention during the past few years is Jackiw-Teitelboim (JT) gravity \cite{Jackiw:1984je,Teitelboim:1983ux}, a model of 2d asymptotically AdS$_2$ dilaton gravity with action:
\begin{equation}
\label{JTaction}
S_{\text{JT}}[g,\Phi] = \frac{1}{16\pi G_2}\int d^2x \sqrt{-g}\Phi \left(R-\Lambda\right) + S_{\text{GH}},
\end{equation}
defined in terms of the 2d metric $g_{\mu\nu}$ and the dilaton field $\Phi$, with cosmological constant $\Lambda= -2/L^2$. The model is topological in the bulk, but it has non-trivial dynamics due to the choice of boundary conditions. This happens in a very similar way as in the Chern-Simons / WZW correspondence.
\\~\\
Upon path integrating $\Phi$, one finds $R = \Lambda$ and the only physical gravitational degree of freedom in JT gravity is the boundary reparametrization $F(\tau)$ \cite{Almheiri:2014cka,Jensen:2016pah,Maldacena:2016upp,ads2}, whose dynamics is governed by the Schwarzian action, arising entirely from the Gibbons-Hawking boundary term of \eqref{JTaction}:
\begin{equation}
\label{SSch}
S[f] = -C\int d\tau \, \left\{F,\tau\right\}, \qquad \left\{F,\tau\right\} \equiv \frac{F'''}{F'} - \frac{3}{2}\left(\frac{F''}{F'}\right)^2,
\end{equation}
which first appeared describing the low-energy dynamics of Sachdev-Ye-Kitaev (SYK) models (see e.g. \cite{KitaevTalks,Sachdev:1992fk,Polchinski:2016xgd,Jevicki:2016bwu,Maldacena:2016hyu,Jevicki:2016ito,Cotler:2016fpe}). The reparametrization $F(\tau)$ represents the AdS$_2$ Poincar\'e time $F$ in terms of some proper time $\tau$. This means the quantum gravity path integral only contains inequivalent frames in a fixed AdS$_2$ ambient space, and is hence far more tractable than expected generically. 
The finite temperature theory is found by further reparametrizing $F \equiv \tan \frac{\pi}{\beta} f$, in terms of the diff($S^1$) reparametrization $f$, satisfying $\dot{f}\geq0$ and $f(\tau+\beta)=f(\tau)+\beta$. The quantum gravitational thermal correlation functions are then found by computing the thermal Schwarzian path integral
\begin{equation}
\label{pischw}
\left\langle \mathcal{O}\right\rangle_{\beta} \equiv \int_\mathcal{M} \left[\mathcal{D}f\right]\mathcal{O}[f] e^{C \int_{0}^{\beta}d\tau \left\{\tan \frac{\pi}{\beta} f,\tau\right\}},
\end{equation}
with a suitable operator insertion $\mathcal{O}[f]$, and with integration space $\mathcal{M}=$ diff($S^1$)/SL(2,$\mathbb{R}$). In the semi-classical regime (at large $C$), this path integral localizes to its classical equation of motion with solution $f(\tau) =  \tau$. Such path integrals are always performed in Euclidean signature, with the resulting expressions (carefully) Wick-rotated afterwards to real time \cite{Mertens:2017mtv}.
\\~\\
Given any (real-time) off-shell reparametrization map $F(t)$ on the boundary curve of a patch of AdS$_2$, we can set up a unique bulk frame by shooting in (and extracting) null rays at times $t-z$ and $t+z$. This, combined with conformal gauge, uniquely defines a bulk point $(t,z)$, and constructs the bulk metric:
\begin{equation}
\label{bulkmetric}
ds^2 = \frac{F'(u)F'(v)}{(F(u)-F(v))^2}(d z^2-d t^2),
\end{equation}
where the conformal factor is hence determined by the construction. This fully fixes the (small) bulk diffeomorphism gauge invariance. We will use this particular choice of bulk coordinates to define the bulk observables. This definition relates everything to boundary-intrinsic operations, in effect anchoring the definition of bulk observables to the holographic boundary line. The latter plays the role of a reference platform \cite{Donnelly:2015hta,Giddings:2018umg}. We have previously explored this definition of bulk coordinates in \cite{Blommaert:2019hjr}. 
\\~\\
In this note, we present three separate computations that can be done within this model that address aspects of the Hawking evaporation process. The computations are logically distinct and illustrate the power of the JT model to investigate these problems. 
Section \ref{s:Unruh} applies the above bulk frame to the bulk matter stress tensor and the matter occupation numbers in the Unruh heat bath surrounding the black hole, i.e. we construct the generalization of Unruh's effect including matter-gravitational interactions. Section \ref{sect:matent} discusses the entanglement entropy of matter fields in AdS$_2$ across a bulk entangling surface, emphasizing an invariant definition of the location of the surface. Section \ref{sect:evap} proceeds within the standard semi-classical framework, but includes evaporating boundary conditions. The evaporating black hole in JT gravity was constructed in \cite{ads2}. Here we extend this study and compute the entanglement of outgoing matter (early radiation) with the remaining interior (late radiation) and obtain an analytic result. The result demonstates information loss within the semi-classical framework quite explicitly. \\
Of course, ultimately we would want to perform an exact quantum gravitational computation of the evaporating black hole. Initial steps in this direction were taken in \cite{Almheiri:2018xdw}, but a full understanding is still lacking. We leave this to future work. 
\\~\\
For later reference, we write down the Schwarzian partition function \cite{Maldacena:2016upp,Cotler:2016fpe,Stanford:2017thb}:
\begin{equation}
Z = \left\langle \mathbf{1}\right\rangle_{\beta} = \left(\frac{2\pi C}{\beta}\right)^{3/2}e^{\frac{2\pi^2 C}{\beta}},
\end{equation}
the Schwarzian derivative expectation value \cite{Stanford:2017thb,Mertens:2017mtv}:
\begin{equation}
\label{sch1pt}
\left\langle \left\{\tan \frac{\pi}{\beta}f , \tau\right\}\right\rangle_{\beta} = \frac{1}{\beta Z} \frac{\partial Z}{\partial C} = \frac{2\pi^2}{\beta^2} + \frac{3}{2C \beta},
\end{equation}
and the (Euclidean signature) bilocal correlation function \cite{altland,altland2,Mertens:2017mtv,Mertens:2018fds,paper3,Blommaert:2018iqz,kitaevsuh,zhenbin}:
\begin{align}
\label{sch2pt}
G^{\beta}_{\ell,C}(\tau_1,\tau_2)  &= \left\langle \mathcal{O}_\ell(\tau_1,\tau_2)\right\rangle_\beta = \left\langle \left(\frac{f'_1f'_2}{\frac{\beta^2}{\pi^2}\sin \frac{\pi}{\beta}(f_1-f_2)^2}\right)^\ell\right\rangle_{\beta} \nonumber\\
 &= \frac{4}{(2C)^{2\ell}Z} \int d\mu(k_1) \int d\mu(k_2)\, e^{ - \tau \frac{k_1^2}{2C}}\,  e^{-(\beta-\tau) \frac{k_2^2}{2C}}\frac{ \Gamma( \ell \pm i k_1 \pm i k_2)}{{2\pi^2}\, \Gamma(2\ell)},
\end{align}
where $d\mu(k)= dk k \sinh(2\pi k)$ and the $\pm$-notation means taking the product of all cases.

\section{Unruh heat bath}
\label{s:Unruh}
We couple the JT model \eqref{JTaction} to the free boson action:
\begin{equation}
S_{\text{mat}} = \frac{1}{2} \int d^2x \sqrt{-g}(\partial_\mu \phi)^2,
\end{equation}
and want to study the effect of particle creation in different matter vacuum states \cite{Spradlin:1999bn}, labeled by the reparametrization functions $f(u)$ and $f(v)$ in the bulk geometry \eqref{bulkmetric}. Considering the thermal state, we will write down the Unruh energy fluxes in this system, and decompose these to read off how the Planckian black body spectrum and the thermal atmosphere are modified in quantum gravity for the eternal black hole.

\subsection{Energy flux}
\begin{figure}[H]
\centering
\includegraphics[width=0.7\textwidth]{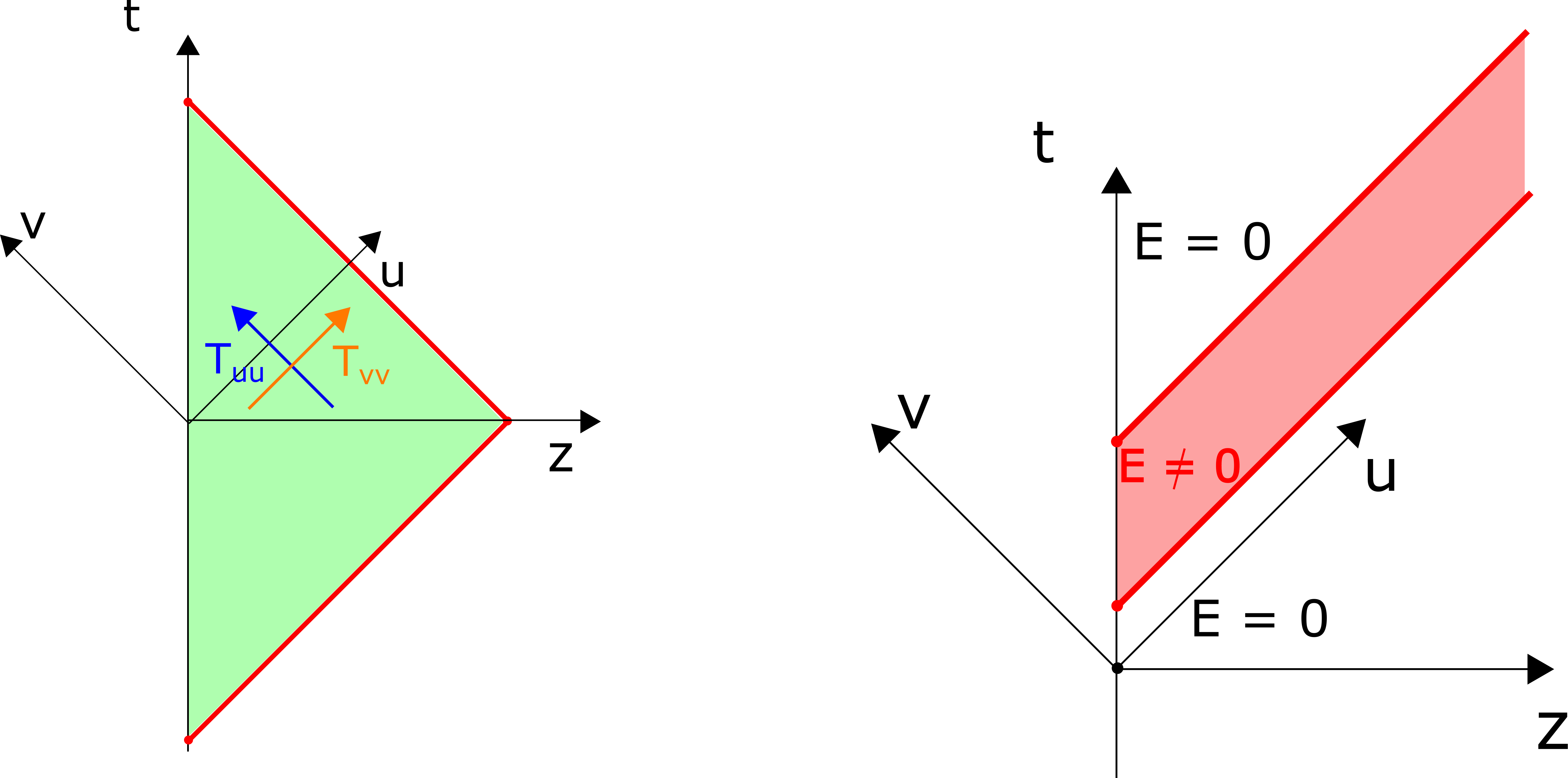}
\caption{Left: Local bulk fluxes of energy. Outgoing flux $T_{uu}$ and ingoing flux $T_{vv}$. Right: Bilocal boundary operation insertion at zero temperature, and the bulk injection of energy that it entails. The bulk energy densities $T_{uu}$ and $T_{vv}$ are zero before and after the injections, and non-zero but constant (in momentum space) and equal in between the ends of the bilocal. In the semi-classical regime, a fixed energy $E(\ell,t_{12})$ is injected by these operators. }
\label{Unruh}
\end{figure}

Within this matter theory, for a fixed background $F$, the propagator is well-known:\footnote{\label{fn1}This formula implicitly contains a $1/Z$ factor in the lhs. Since the partition function $Z$ in a 2d CFT depends on the background metric $F$ through the conformal anomaly, one might be worried about its role in this computation. However, it was recently shown in appendix C of \cite{zhenbin} that the conformal anomaly only causes a shift of the $C$-coefficient of the Schwarzian that is moreover subdominant to the $C$ coming from the gravitational sector.}
\begin{align}
\left\langle \phi(u,v) \phi(u',v')\right\rangle_{\text{CFT}} 
&= -\frac{1}{4\pi}\ln \abs{\frac{(F(u)-F(u'))(F(v)-F(v'))}{(F(v)-F(u'))(F(u)-F(v'))}} \label{greenslc},
\end{align}
which is the 2d CFT two-point function supplemented with an image charge term such that Dirichlet boundary conditions are specified at the holographic boundary $z=0$. 
The stress tensor components are given by
\begin{equation}
\label{stresscompo}
T_{uu} = \partial_u \phi \partial_u \phi, \qquad T_{vv} = \partial_v\phi \partial_v\phi,
\end{equation}
interpreted as the outgoing and ingoing energy densities (Figure \ref{Unruh} left).

\noindent As composite operators in the quantum theory, these require renormalization. Referring w.r.t. the Poincar\'e frame and using \eqref{greenslc} with a point-splitting regularization, we find the renormalized operators:
\begin{align}
\left\langle :T_{uu}(u):\right\rangle_{\text{CFT}} &= -\frac{1}{4\pi}\lim_{u'\to u}\left[\frac{F'(u)F'(u')}{(F(u)-F(u'))^2} - \frac{1}{(u-u')^2}\right], \\
\left\langle :T_{vv}(v):\right\rangle_{\text{CFT}} &= -\frac{1}{4\pi}\lim_{v'\to v}\left[\frac{F'(v)F'(v')}{(F(v)-F(v'))^2} - \frac{1}{(v-v')^2}\right].
\end{align}
At finite temperature we set $F \equiv \tanh\frac{\pi}{\beta}f$ in terms of the reparametrization $f$. This is the generalization of the parametrization mentioned in the Introduction in Euclidean time. \\
 Series expanding the above expression, one rewrites these as\footnote{The CFT expectation value brackets are left implicit from here on.}
\begin{equation}
\label{Unruhflux}
:T_{uu}(u): = -\frac{c}{24\pi}\left\{\tanh\frac{\pi}{\beta}f(u),u\right\}, \qquad :T_{vv}(v): = -\frac{c}{24\pi}\left\{\tanh\frac{\pi}{\beta}f(v),v\right\},
\end{equation}
where we have introduced the matter central charge $c$ to generalize to a generic matter CFT sector. The boundary time frame is extrapolated into the bulk using these equations, in the light-ray fashion described in the Introduction.\footnote{These operators are of the type of \eqref{pischw}, and should be written as $:T_{uu}[f(u)]:$ and $:T_{vv}[f(v)]:$, which we won't do to avoid cluttering the equations. Note that these stress tensor components can be considered as diff-invariant observables the way we constructed them, in spite of their tensor indices. Indeed, the tensor indices are found in a boundary-intrinsic way by taking the derivatives in \eqref{stresscompo} in a limiting procedure, a $u$-derivative varies the final boundary point, and a $v$-derivative varies the initial boundary point of the radar definition of the bulk point $(u,v)$.} \\
In effect, bulk energy densities can be computed by inserting the boundary energy operator $T_{tt}(t) = -C \left\{F(t),t\right\}$ (see e.g. \eqref{stressbdy}), up to some prefactors, and reinterpreting the time $t$ as either $u$ or $v$.\footnote{Note the presence of the factor of $C$ in the boundary energy. This has dimensions of length, and hence indeed the bulk stress tensor has dimension $L^{-2}$, and the boundary stress tensor has units of  $L^{-1}$.} This bulk gauge is a choice, but it is one that nicely contains the semi-classical Unruh physics as we illustrate now.
\\~\\
On the saddle $f(t) = t$, and $\left\{\tanh\frac{\pi}{\beta}f(u),u\right\} = -\frac{2\pi^2}{\beta^2}$, leading to the Unruh heat bath \cite{Spradlin:1999bn}
\begin{equation}
\label{Enmat}
\normal{T_{uu}(u)} = \normal{T_{vv}(v)} =  c\frac{\pi}{12}T_H^2,
\end{equation}
in terms of the Hawking temperature $T_H$ of the black hole. Performing instead the full Schwarzian path-integral \eqref{pischw} for a temperature $\beta^{-1} \equiv T_H$, we find using \eqref{sch1pt}:\footnote{The conformal anomaly determines the remaining stress tensor component to be $\left\langle T_{uv}\right\rangle =  \frac{c}{6 L^2}$, in terms of the AdS length $L$. This also holds when doing the full path integral \eqref{pischw}. This is independent of the temperature and can be viewed as an energy offset $E_0$.}
\begin{equation}
\label{exUnruh}
\left\langle \normaltwo{T_{uu}(u)}\right\rangle_{\beta} \,=\, \left\langle \normaltwo{T_{vv}(v)}\right\rangle_{\beta} =  c\frac{\pi}{12}T_H^2 + c\frac{T_H}{16\pi C},
\end{equation}
which includes a further quantum thermodynamical correction that is suppressed in the semi-classical regime $C\to + \infty$.\footnote{By inverse Laplace transforming, for a pure energy quantum state $\left|M\right\rangle$, with energy $M/2C$, we obtain instead
\begin{equation}
\label{Enps}
\left\langle \normaltwo{T_{uu}(u)}\right\rangle_{\text{M}} = \left\langle \normaltwo{T_{vv}(v)}\right\rangle_{\text{M}}=  c\frac{M}{48\pi C^2}.
\end{equation}
}
The measured stress tensor components of observers whose detectors are calibrated to the $(u,v)$-vacuum, is hence spacetime-independent. This changes when matter is being injected (or extracted) into the system. 

\subsection{Energy pulses}
\label{sec:pulse}
Classically, energy can be injected through pulses as studied extensively in \cite{Almheiri:2014cka,ads2}. At the quantum level, this can be done by using bilocal operators of the type in \eqref{sch2pt} (Figure \ref{Unruh} right). Let us explain this relation in more detail.
\\~\\
In Appendix \ref{app:timed}, we demonstrate the Ward identity for boundary stress tensor insertions $T_{tt}(t)$ in bilocal correlators. Dropping contact terms, and continuing to Lorentzian signature, we write:
\begin{align}
\left\langle \mathcal{T}\, T_{tt}(t) \mathcal{O}(t_1,t_2)\right\rangle_\beta \, = \, \Big(\theta(t_1<t<t_2)i\partial_{t_{12}} - \partial_\beta \Big) G^\beta_{\ell,C}(t_1,t_2) + \text{(contact)},
\end{align}
in momentum space interpreted as energy $\frac{k_1^2}{2C}$ between the ends of the bilocal, and $\frac{k_2^2}{2C}$ outside. In particular, one finds energy is conserved everywhere except at the bilocal points:
\begin{align}
\label{encons}
\left\langle \partial_t T_{tt}(t) \mathcal{O}(t_1,t_2)\right\rangle_\beta 
\, = \, \Big( \delta(t-t_1)i\partial_{t_{12}} - \delta(t-t_2)i\partial_{t_{12}}\Big)G^\beta_{\ell,C}(t_1,t_2) + \text{(contact)},
\end{align}
interpreted in Fourier space as injecting and extracting an energy $\frac{k_1^2-k_2^2}{2C}$ at the bilocal points. Since $\normal{T_{uu}(u)}$ and $\normal{T_{vv}(v)}$ are also given by Schwarzian derivatives \eqref{Unruhflux}, the result \eqref{encons} implies there are no transient phenomena for these bulk stress tensors after crossing energy pulses (Figure \ref{Unruh} right).
\\~\\
Let us prove that this operator indeed gives the correct semi-classical energy pulse interpretation. In the limit where we take $N$ bilocals of $\ell=1$ at the same endpoints and with $\ell N \sim C \to + \infty$ to reach the semi-classical regime, the bulk interpretation is a semi-classical coherent state: a null pulse ($m^2=\ell(\ell-1)=0$) of energy $E(\ell N,t_{12})$ that can be found by solving a transcendental equation \cite{Lam:2018pvp}, followed by a negative energy null pulse with energy $-E(\ell N,t_{12})$. One finds the bulk energy densities \eqref{Enps} in between the null pulses, with  $M = E(\ell N, t _{12})$, and zero outside. The classical time reparametrization profile is
\begin{align}
f(t) = \left\{
                \begin{array}{ll} \, t-t_1, \quad t<t_1, \nonumber \\
\, \frac{\sqrt{\frac{2C}{E}}\tanh\left(\sqrt{\frac{E}{2C}}(t-t_1)\right)}{\frac{\ell}{2C}\sqrt{\frac{2C}{E}}\tanh\left(\sqrt{\frac{E}{2C}}(t-t_1)\right)+1}, \quad t_1 <t <t_2, \label{zerotsolution}\\
\, t - t_2 + f_2 , \quad t_2 < t, \nonumber
\end{array} \right.
\end{align}
and represents the Poincar\'e frame, going through a thermal phase, and then returning to a (Shapiro time-delayed) Poincar\'e frame.
\\~\\
The classical problem of measuring the ingoing and outgoing energy densities $T_{vv}$ and $T_{uu}$ within the different regions of Figure \ref{Unruh} right, is translated in the full quantum theory in the operator ordering of the stress tensor w.r.t. the bilocal.\footnote{One can also place it in between both ends of the bilocal operator.} For instance, $\left\langle \mathcal{O}_\ell(t_1,t_2) T_{uu}(u)\right\rangle_{\beta \to + \infty}$ always vanishes, irrespective on the value of $u$ w.r.t. $t_1$ and $t_2$. The amplitude can be read as taking empty AdS, time-evolving to $t=u$, applying $T_{uu}$, then evolving back or forward to the first bilocal time and inject energy, evolve to the second bilocal time and extract the energy, to arrive at empty AdS again. This is similar to the set-ups of \cite{Almheiri:2017fbd,Shenker:2013yza}. \\
The explicit link of the bilocal operator insertions to semi-classical gravitational shockwaves was made in \cite{Lam:2018pvp}, in the context of an out-of-time-ordered four-point function in the Schwarzian model. Within such an OTO-four-point function, one can apply $T_{uu}$ in any ordering w.r.t. the four operators, semi-classically measuring the energy density in each of the sectors of the shockwave diagram.

\subsection{Planckian black body spectrum}
\label{sect:bbspec}
\begin{figure}[h]
\centering
\includegraphics[width=0.55\textwidth]{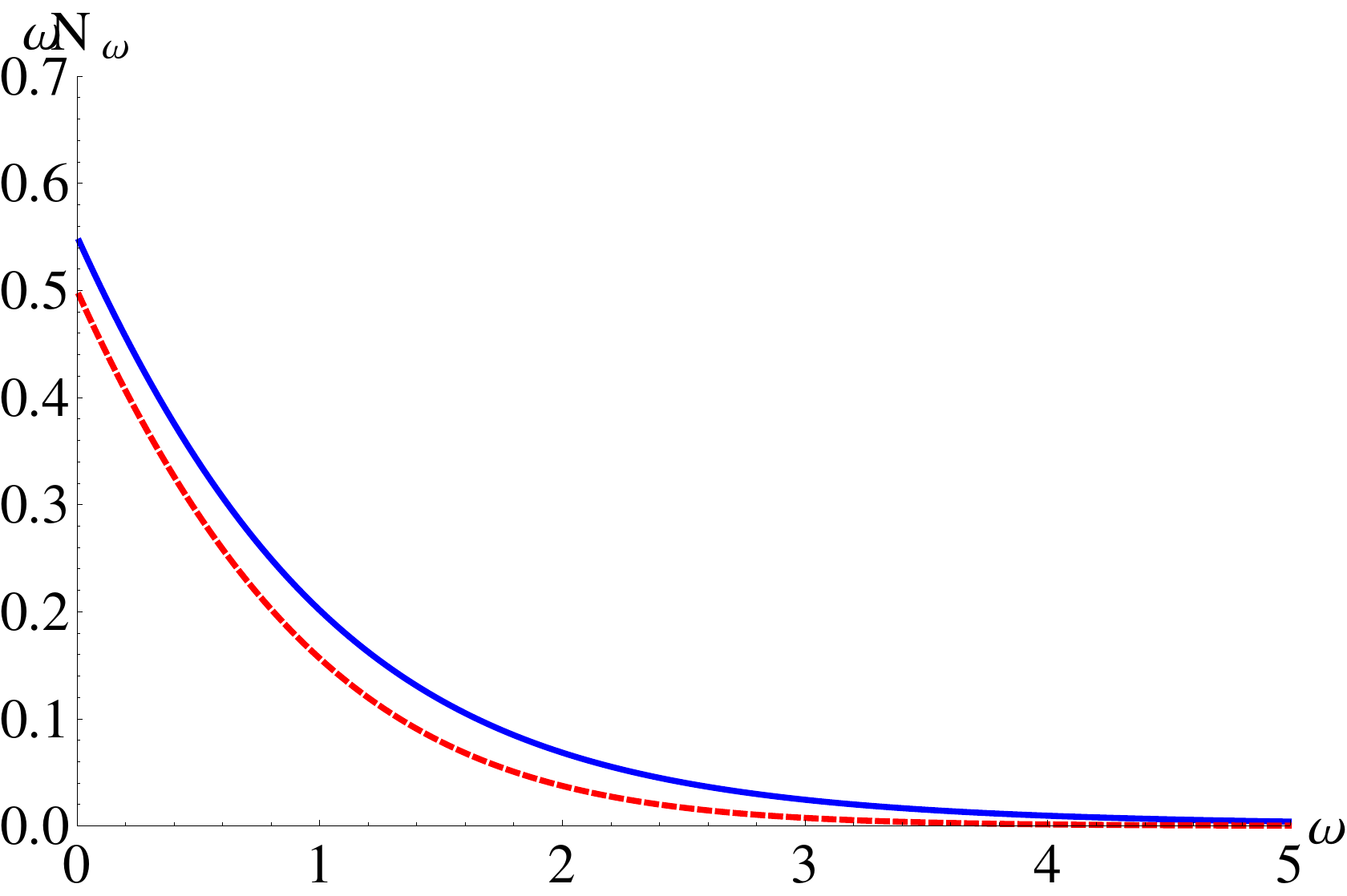}
\caption{Blue (upper): exact energy spectral density $\omega \left\langle N_\omega \right\rangle_{\beta}$ of the Unruh radiation, computed from \eqref{exactplanck}with $\beta=2$. Red (lower): semi-classical Planck black body spectrum of Unruh radiation, coming from \eqref{semiplanck}.}
\label{PlanckE}
\end{figure}

\begin{figure}[h]
\centering
\includegraphics[width=0.55\textwidth]{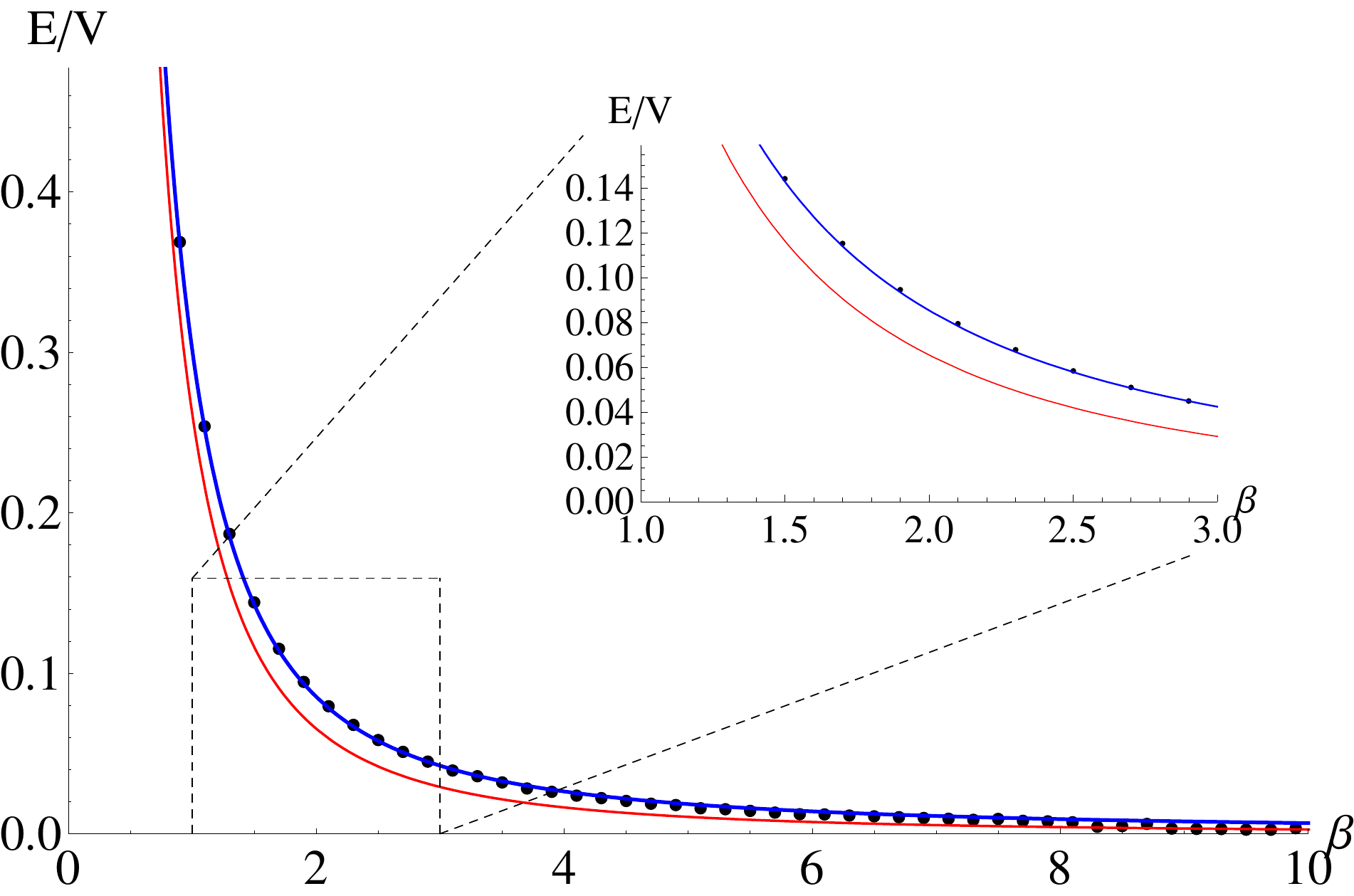}
\caption{Total energy density $\frac{1}{V} \int_{0}^{+\infty}d\omega \, \omega \left\langle N_\omega \right\rangle_{\beta}$ of the Unruh radiation, as a function of $\beta$, computed by integrating \eqref{exactplanck} (black dots). The exact energy \eqref{exUnruh} (with $c=1$) is plotted as a blue line (top). The semi-classical energy \eqref{Enmat} is plotted as a red line (bottom), computable by integrating \eqref{semiplanck}. The inset shows in more detail the match at the exact level, and the approximation made by taking the semi-classical result.}
\label{Plancknum}
\end{figure}

Next we perform a spectral decomposition of the Unruh fluxes \eqref{exUnruh}. It is well-known, using 2d CFT techniques, that the occupation number of the chiral mode $u_\omega(y) = \frac{1}{\sqrt{4\pi \omega}}e^{-i\omega y}$, i.e. a positive frequency mode in the observer's local frame $y$, written as $N_\omega[f] \equiv \left\langle 0_F\right|a^\dagger_\omega a_\omega \left|0_F\right\rangle$ in the vacuum associated to the Poincar\'e frame $F(y) \equiv \tanh \frac{\pi}{\beta}f(y)$, is given by \cite{Fabbri:2004yy}:
\begin{equation}
\label{planck}
N_\omega[f] = -\frac{1}{\pi}\int dy_1 \int dy_2 u_\omega(y_1) u^*_\omega(y_2) \left[\frac{f'(y_1)f'(y_2)}{\frac{\beta^2}{\pi^2}\sinh^2 \frac{\pi}{\beta}(f_1-f_2)} - \left(\frac{1}{y_{12}}\right)^2\right].
\end{equation}
The integral is finite at $y_1=y_2$ due to the renormalization w.r.t. the reference state $\left|0_y\right\rangle$. In these formulas, $y$ can be either $u$ or $v$.
\\~\\
Semiclassically, where $f(y)=y$, the treatment is well-known and leads to the Planckian black body spectrum. Let us briefly write down how this is proven. Firstly, since the integrand only depends on $y_1-y_2$ one of the integrals factorizes out and gives a divergent prefactor $2\pi\delta(0)$.\footnote{Due to the reflecting boundary conditions at $z=0$, the field $\phi(u,v) = \phi_L(v) + \phi_R(u)$ has mode oscillators related by $a_{R\omega} = -a_{L\omega}$. This means there is only one set of oscillators and left- and right-moving modes are related by the doubling trick. The integral over $y_1+y_2$ mentioned here ranges from $-\infty$ to $+\infty$ \emph{after} mirror doubling the system. Note also that one is free to choose $y_1$ and $y_2$ to be either $u$ or $v$ independently without changing the result. This would select one of the four terms of \eqref{greenslc} to relate the occupation number to $\partial_{\pm}\phi \partial_{\pm}\phi$, all of which yield the same outcome.} For the remaining integral one can use the Fourier transform formula
\begin{align}
\label{fouri}
-\frac{1}{2\pi\omega}\int_{-\infty}^{+\infty} dt \left(\frac{1}{\frac{\beta}{\pi}\sinh(\frac{\pi}{\beta}(t \mp i \epsilon))}\right)^{2}e^{-i\omega t} &= \frac{1}{\beta\omega} e^{\mp\frac{\beta}{2} \omega}\Gamma\left(1 + i \frac{\beta}{2\pi} \omega\right) \Gamma\left(1 - i \frac{\beta}{2\pi} \omega\right) \nonumber \\
&= \frac{e^{\mp \frac{\beta}{2}\omega}}{e^{\frac{\beta}{2}\omega}-e^{-\frac{\beta}{2}\omega}}.
\end{align}
In the limit $\beta \to +\infty$, we obtain the formula:
\begin{equation}
\label{PCref}
-\frac{1}{2\pi\omega}\int_{-\infty}^{+\infty} dt \frac{1}{(t \mp i \epsilon)^2}e^{-i\omega t} = \lim_{\beta  \to + \infty} \frac{e^{\mp\frac{\beta}{2}\omega}}{e^{\frac{\beta}{2}\omega}-e^{-\frac{\beta}{2}\omega}} =  \mp\Theta(\mp\omega),
\end{equation}
which is readily found from the integral definition of the Heaviside function. Both of these formulas are regularized separately by moving the pole at $t=0$ in the same direction. In most treatments, one first subtracts these terms before performing the Fourier transform, rendering the regularization $t\to t \mp i\epsilon$ obsolete.
\noindent Hence subtracting out the vacuum contribution \eqref{PCref}, one obtains the Planckian spectrum:
\begin{equation}
\label{semiplanck}
N_\omega[f(y)=y] \,=\,  \frac{e^{-\frac{\beta}{2}\omega}}{e^{\frac{\beta}{2}\omega}-e^{-\frac{\beta}{2}\omega}}.
\end{equation}
Beyond semi-classical gravity, the two choices $\mp i \epsilon$ are inequivalent, and we define the correct quantum operator to be the \emph{average} of time-ordered ($+i\epsilon$) and anti-time-ordered ($-i\epsilon$) correlation functions, effectively the Hadamard two-point function.\footnote{In fact, considering the integral $\int_{-\infty}^{+\infty} d\omega \, \omega \times \, \eqref{planck}$, one needs to use $\omega \to - \omega$ and $t \to -t$ transformations to map this into an integral over positive $\omega$ only and identify it with the positive energy spectral density. These transformations then immediately correspond to taking the average of time-ordered and anti-time-ordered correlators.}
\footnote{A similar conclusion was found for the bulk quantum metric in \cite{Blommaert:2019hjr}, in that case by insisting on hermiticity of the metric operator.} We hence path-integrate over the frame $f$ as in \eqref{pischw}:
\begin{equation}
\big\langle N_\omega \big\rangle_{\beta} = \int \left[\mathcal{D}f\right]N_\omega[f] e^{C \int_{0}^{\beta}d\tau\left\{\tan \frac{\pi}{\beta}f,\tau\right\}},
\end{equation}
using the average of both operator orderings for \eqref{planck}. Using the fact that the bilocal operator commutes with the Hamiltonian of the Schwarzian system \cite{Mertens:2017mtv}: 
\begin{equation}
\left[\frac{f'(y_1)f'(y_2)}{ \frac{\beta^2}{\pi^2}\sinh^2 \frac{\pi}{\beta}(f_1-f_2)}, H_{\text{Schw}}\right]=0,
\end{equation}
the result again only depends on the difference $y_{12}$.\footnote{Note that this need not occur for each single path $f$ in the integral.} This means the result is independent of $y_1+y_2$, and the integral gives $2\pi \delta(0)$, the same universal divergence also appearing at the semi-classical level. Using the known expressions for the Schwarzian bilocal for $\ell=1$ and averaging time-ordered and anti-time-ordered expressions, we write:\footnote{We set $C=1/2$ here.}
\begin{align}
\big\langle N_\omega \big\rangle_{\beta} + \frac{1}{2} = \frac{1}{2\pi^3\omega}\frac{1}{Z}\int_{-\infty}^{+\infty} dt e^{-i\omega t} \int &d\mu(k_1)d\mu(k_2) \Gamma(1 \pm i k_1 \pm i k_2) \nonumber \\
&\times \left(e^{- i t (k_1^2-k_2^2)} + e^{+ i t (k_1^2-k_2^2)} \right) e^{-\beta k_2^2}
\end{align}
the $+ \frac{1}{2}$ coming from \eqref{PCref}. Performing the $t$-integral yields delta-functions $\delta(\omega \pm k_1^2 \mp k_2^2)$. We get:
\begin{equation}
\label{exactplanck}
\big\langle N_\omega \big\rangle_{\beta} = \frac{1+e^{-\beta \omega}}{2\pi^2 \omega} \frac{1}{Z}\int d\mu(k) \sinh(2\pi \sqrt{\omega+k^2})e^{-\beta k^2}\Gamma(1 \pm i k \pm i \sqrt{\omega + k^2}) - \frac{1}{2}.
\end{equation}
This integral can be done numerically and is plotted in Figure \ref{PlanckE}.
\\~\\
Integrating the energy spectral density over $\omega$,  one obtains\footnote{The doubling trick is also performed here.}
\begin{equation}
\label{checkk}
\int_{0}^{+\infty} d\omega \, \omega \, \big\langle  N_\omega \big\rangle_{\beta} = \int_{-\infty}^{+\infty} dy \left\langle \hspace{-0.15cm}\normal{T_{\pm\pm}}\hspace{-0.15cm}\right\rangle_\beta =  \int_0^{+\infty} dy \left\langle\sqrt{-g}g^{00} \hspace{-0.15cm}\normal{T_{00}}\hspace{-0.15cm}\right\rangle_\beta,
\end{equation}
where in the last equality we used that $\sqrt{-g}g^{00} =1$, also off-shell. Indeed, the quantum-corrected Unruh population \eqref{exUnruh} is slightly more energetic than the semi-classical one \eqref{Enmat}, leading to a larger population of the thermal modes as Figure \ref{PlanckE} shows. \\
The $y$-integral in \eqref{checkk} factorizes and can be identified as $2\pi \delta(0) = V$. We have checked numerically that indeed the quantum term in \eqref{exUnruh} is found by numerically doing the $\omega$-integral of \eqref{checkk} (Figure \ref{Plancknum}).\footnote{The formula of v1 of this paper did not average over time-ordered and anti-time-ordered correlators, and did not pass this consistency check.}
\\~\\
The radiation is not precisely thermal. The UV region $(\omega \gg 1$) is dominated by the $\tau \to 0$ pole and is the same as for the semi-classical Planck spectrum, but deviations due to gravitational interactions are visible at lower energies. This means quantum gravitational effects modify the Unruh process. This also means that there is information stored within the heat bath of the black hole, but it is not visible at the semi-classical level. It is well-known that matter interactions do not influence the thermal character of the Unruh effect \cite{Gibbons:1976es}, basically because one can do perturbation theory and one finds thermal answers at every fixed order. The non-thermality hence fully originates from the matter-gravitational interactions. This is to be expected, and was also observed for the metric tensor itself in \cite{Blommaert:2019hjr}. \\
It is remarkable that we are able to obtain an analytic formula describing the exact quantum gravitational Unruh spectrum.
\\~\\
The zero-temperature limit of \eqref{exactplanck} gives:
\begin{equation}
\big\langle N_\omega \big\rangle_{\beta\to\infty} = \lim_{\beta \to \infty} \frac{1+e^{-\beta \omega}}{4\pi^2 \omega}\sinh(2\pi \sqrt{\omega})\Gamma(1 \pm i \sqrt{\omega})  - \frac{1}{2} = \frac{1}{2}\left(\coth \pi \sqrt{\omega} - 1\right).
\end{equation}
This still contains an interesting lesson. At zero temperature, the vacuum becomes $\left|0_F\right\rangle \to \left|0_f\right\rangle$, instead of the local observer's vacuum $\left|0_y\right\rangle$. And it still contains low-energy particles $\omega \ll 1$. 
\\~\\
Starting with \eqref{exactplanck}, the semi-classical regime where one reproduces the standard Unruh results, is when $\omega \ll k^2$, where $k$ is evaluated at its saddle point. This ensures the number operator insertion does not backreact on the geometry (influencing the location of the saddle), and is really a semi-classical measurement. The zero-temperature result becomes semi-classical when $\omega \gg 1$.
\\~\\
Starting with these expressions, one can readily find related operators that can be used to create and remove particles. E.g. using $\big[a_\omega,a^\dagger_{\omega'}\big] = \delta(\omega-\omega')$:
\begin{equation}
\Big\langle \left\langle 0_F\right|a_\omega a^\dagger_\omega \left|0_F\right\rangle\Big\rangle_{\beta\to\infty} = \frac{1}{2}\left(\coth \pi \sqrt{\omega} + 1\right),
\end{equation}
giving the normalization $1$ for the high-energy modes, but deviating from this at lower energies. Since it is constructed from the bilocal Schwarzian operator, one can readily separate the two oscillators, i.e. put other operators in between, and then utilize the Schwarzian diagrammatic rules of \cite{Mertens:2017mtv} to write down the amplitude.
\\~\\
Following \cite{Mertens:2019tcm}, one can isolate a defect insertion of the number operator from \eqref{exactplanck} as:
\begin{equation}
D_{N_\omega}(k) = \frac{1}{4\pi^2 \omega} \sinh(2\pi\sqrt{k^2-\omega}) \Gamma(1 \pm i k \pm i \sqrt{k^2 - \omega}) \Theta(k^2-\omega) + (\omega \to - \omega) - \frac{1}{2},
\end{equation}
applied within the region of the disk with momentum label $k$.
\\~\\
The analytic solvability of the massless free scalar field, readily extends to other matter fields. \\
The extension of the story to charged matter in a charged black hole is instructive, and is presented in appendix \ref{app:charge}. \\
Changing to a single fermionic field instead, one can wonder about how the low-energy quantum gravitational corrections to the occupation level are modified due to Pauli expulsion. This is an interesting calculation, which we perform in appendix \ref{app:fermion}. \\
Another extension one can pursue is that of a massive scalar field. The same technique to derive \eqref{planck} can be used in the massive case to relate the occupation number to $\left\langle \partial_t \phi_1 \partial_t \phi_2\right\rangle^{\text{QFT}}_{\beta}$, where the two-point function $\left\langle \phi_1 \phi_2\right\rangle$ for a massive bulk scalar field is given in terms of a $_2F_1$ hypergeometric function. The procedure to compute its Schwarzian path integral was sketched in \cite{Blommaert:2019hjr}, in principle solving this problem. It remains to be seen whether the resulting expressions can be written down in an illuminating way.
\\~\\
Finally, one can change both the oscillators and the state in \eqref{planck} independently by changing respectively the modes $u_\omega$ and the bilocal operator. These could then be used to analyze the observations by other observers, e.g. in response to energy injections from boundary bilocals as in section \ref{sec:pulse}. We leave a study of the physical significance to future work.

\section{Matter entanglement entropy}
\label{sect:matent}
\begin{figure}[h]
\centering
\includegraphics[width=0.8\textwidth]{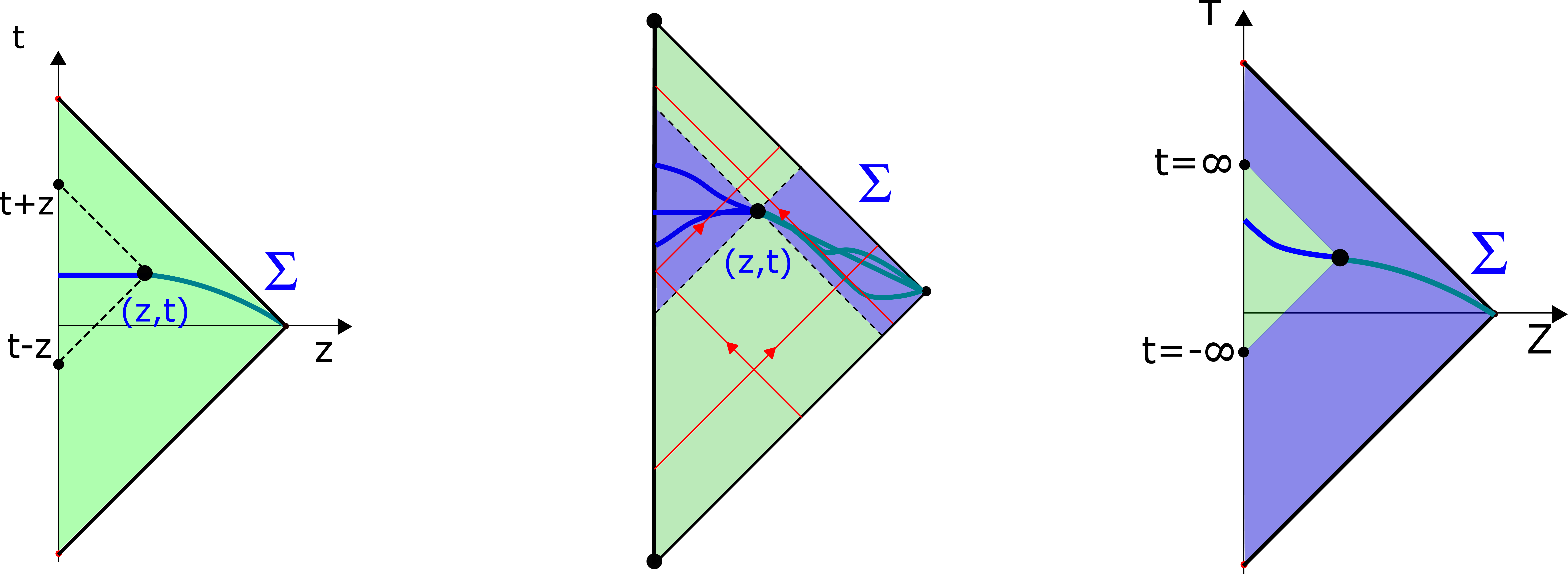}
\caption{Left: Matter entanglement entropy obtained by dividing a Cauchy slice $\Sigma$ in two pieces. The bulk point is at $(z,t)$. Middle: Foliation independence of entanglement entropy on Cauchy slice. One can freely move the Cauchy surface within the blue regions, keeping it spacelike everywhere. Right: $z\to +\infty$ limit, where the entanglement entropy is between the interior (blue) and exterior (green) of the black hole. The full patch is the Poincar\'e frame. The result agrees with the thermal entropy of the CFT gas surrounding the black hole.}
\label{Sentm}
\end{figure}

Next we take a look at the matter CFT entanglement entropy. Divide a spatial slice $\Sigma$ of AdS$_2$ in two parts (Figure \ref{Sentm} left).

\begin{figure}[h]
\centering
\includegraphics[width=0.65\textwidth]{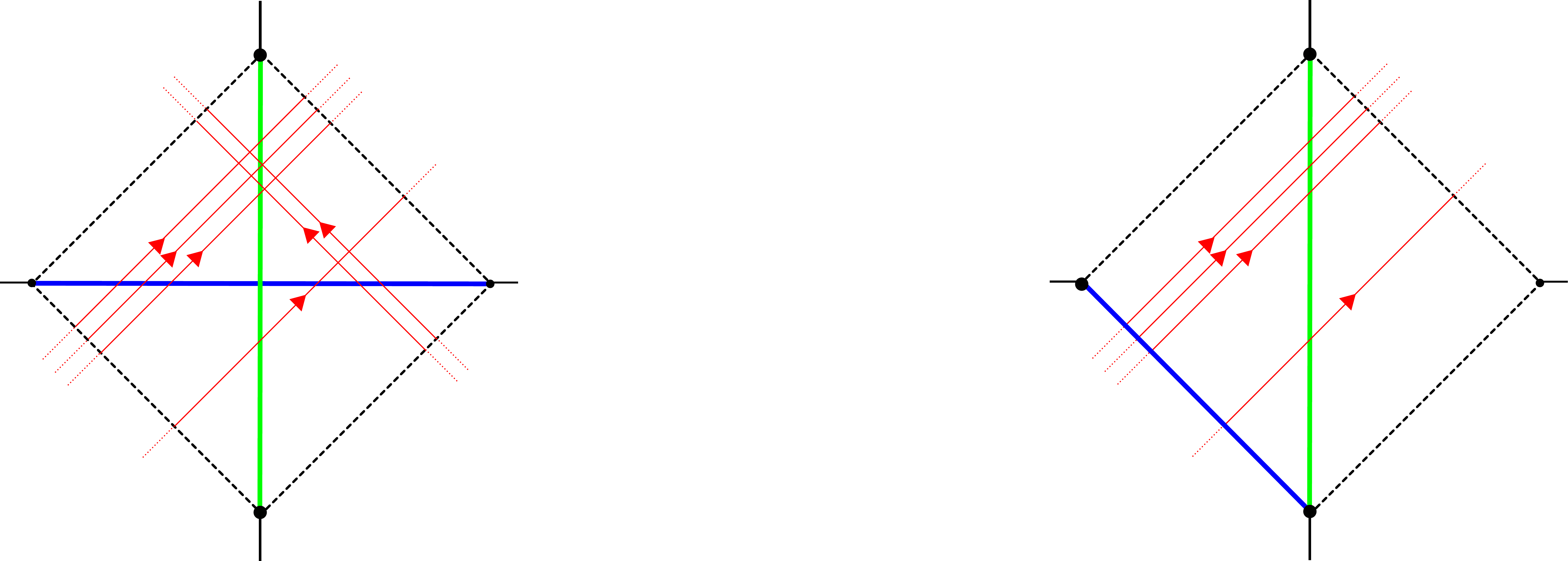}
\caption{Left: In a CFT, one can change a spatial interval (blue) into a time interval (green) using null paths, preserving information flow. Right: In a chiral (sector of a) CFT, one can move the endpoint of the interval (blue) along one of the null directions, e.g. up to the time interval (green), and preserve the information flow.}
\label{SinCFT}
\end{figure}

\begin{figure}[h]
\centering
\includegraphics[width=0.6\textwidth]{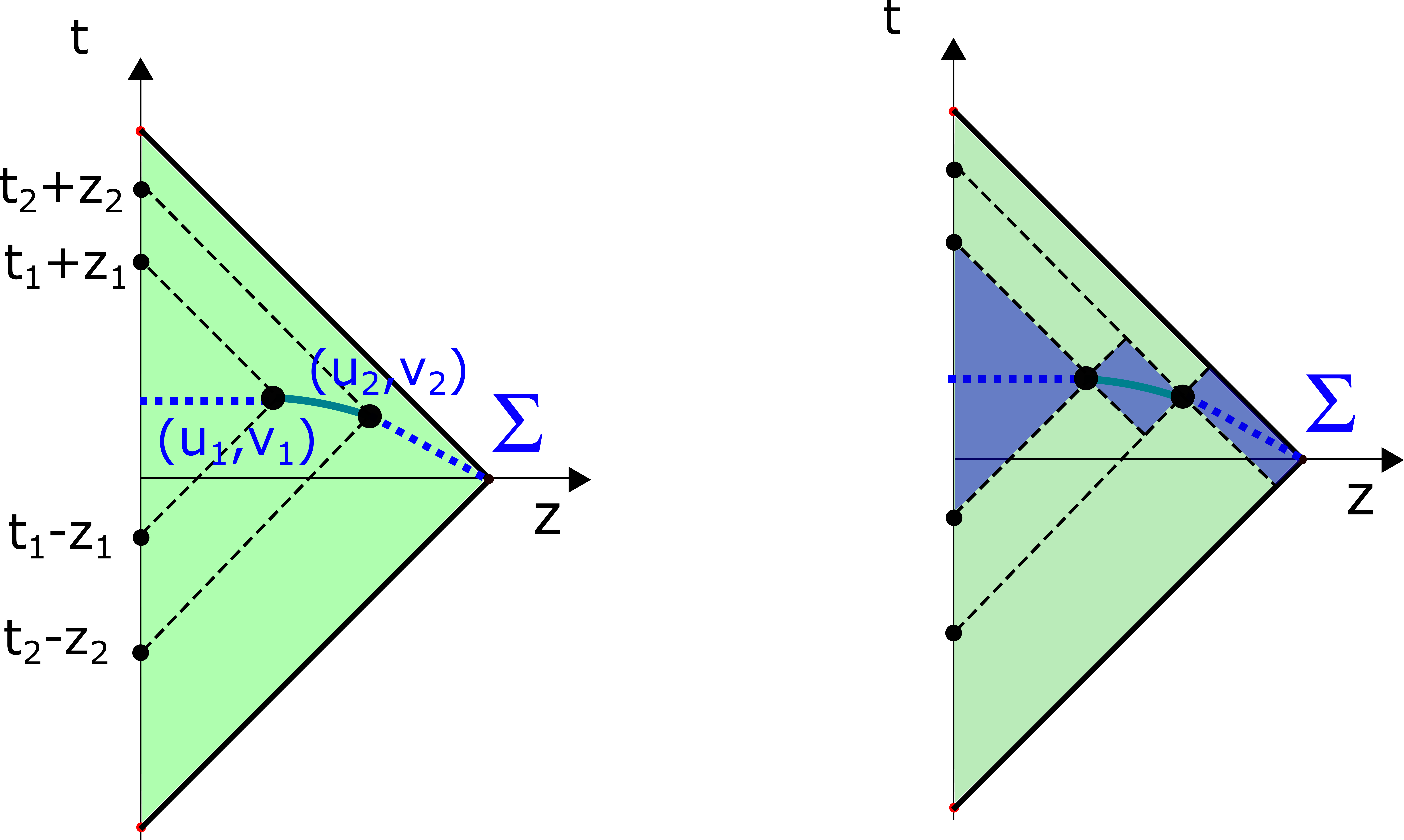}
\caption{Left: Matter entanglement of a bulk interval between $(u_1,v_1)$ and $(u_2,v_2)$). Right: The Cauchy slice $\Sigma$ can be deformed into those new surfaces within the blue regions that respect the spacelike nature.}
\label{Sentbulk}
\end{figure}

\noindent The entanglement entropy between the matter degrees of freedom left and right of the bulk point $u=t+z$, $v=t-z$ of the matter CFT is given by the formula:
\begin{equation}
\label{CFTS}
S_{\text{ent}} = \frac{c}{12} \ln \frac{(f(u)-f(v))^2}{\delta^2 f'(u)f'(v)},
\end{equation}
where the UV cut-off $\delta$ is measured by the observer in the $u,v$-coordinate frame.\footnote{This is half the entanglement entropy of the doubled interval between $(u,v)$ and $(v,u)$. See also \cite{Fiola:1994ir} for a thorough early treatment, and \cite{Callebaut:2018nlq,Callebaut:2018xfu} for a recent analysis within the JT context. We compare the renormalized version of this formula with the general curved space formula of \cite{Fiola:1994ir} in Appendix \ref{sect:enta}.} In our language, using the radar construction of bulk points, this corresponds to a boundary-intrinsic choice of UV-cut-off. This equation excludes gravity, and is taken in the matter Poincar\'e vacuum state, labeled by the coordinate $f$. \\
We do not need to specify the precise shape of the Cauchy surface $\Sigma$ of interest due to the foliation independence of $S_{\text{ent}}$ and the fact that we insist on the UV-cut-off associated to the $u,v$-frame. This is illustrated in Figure \ref{Sentm} middle.\footnote{Different foliations can be obtained by applying conformal transformations $F$ with the fixed point properties $F(f(u))=f(u)$ and $F(f(v))=f(v)$. Additionally, insisting on using the same UV-cutoff yields the same formula. Note that this is different than replacing $f$ in \eqref{CFTS} by $F \circ f$ in both numerator and denominator, which would change the state to the vacuum in the $F$-coordinates.}
\\~\\
Classically, using $f(u)= \tanh \frac{\pi}{\beta}u$, this becomes
\begin{equation}
\label{Ssemi}
S_{\text{ent}} = \frac{c}{6} \ln \frac{\beta}{\pi} \sinh \frac{\pi}{\beta}(u-v) - \frac{c}{6} \ln \delta.
\end{equation}
Setting $z\to + \infty$ to obtain the thermal entropy, we write:
\begin{equation}
\label{Sth}
S_{\text{th}} = \lim_{z\to + \infty} S_{\text{ent}} - S_{\text{ref}} = \frac{c}{6} \frac{2\pi}{\beta} \lim_{z\to+\infty}z,
\end{equation}
which is IR-divergent where $\lim_{z\to+\infty}z \equiv V$, the spatial volume. Note that one might equally well call this a UV horizon divergence, associated to the infinite volume stashed close to the black hole horizon.\footnote{This is similar to the divergence of the thermal gas entropy in Rindler space, for which the UV cutoff in the Rindler radial coordinate at $\rho=\epsilon$ is mapped into a volume divergence in tortoise coordinates at $r = \ln \epsilon \to -\infty$.}  In writing this expression, we subtracted the zero-temperature entropy $S_{\text{ref}} = \frac{c}{6} \ln \frac{u-v}{\delta}$ to isolate the thermal piece \cite{Holzhey:1994we}. This cancels the dependence on the cutoff $\delta$ and gives an additional contribution $\frac{c}{6}\ln \frac{\beta}{4\pi z}$ that is subdominant in the large $z$-regime.
\\~\\
As a check on \eqref{Sth}, the total matter energy is
\begin{equation}
E_{\text{mat}} = -\int dz \sqrt{-g} \hspace{-0.15cm}\normal{T_{0}^{0}} = \int dz \hspace{-0.15cm}\normal{T_{00}} = \frac{c}{6} \frac{\pi}{\beta^2} \lim_{z\to+\infty}z,
\end{equation}
obtained by integrating \eqref{Enmat} over the spatial volume.\footnote{We used $T_{00} = T_{uu}+T_{vv}$. In principle, one should add to this $2T_{uv} = \frac{c}{3L^2}$ from the conformal anomaly and using $R=-2/L^2$ for AdS$_2$ JT-gravity, with $L$ the AdS length. This contribution is $\beta$-independent and hence does not contribute to the entropy.} Using the thermodynamical relation $E_{\text{mat}} = \partial_\beta \beta F$, one finds $F = -E_{\text{mat}}$ and the thermal entropy $S=\beta(E_{\text{mat}}-F)$ is indeed given by \eqref{Sth}. This is the semi-classical result that the thermal entropy of the matter gas surrounding the black hole can be viewed as entanglement entropy of the half-space (Figure \ref{Sentm} right).
\\~\\
The formula \eqref{CFTS} can also be read as the (analytically continued) geodesic length between two boundary points $v$ and $u$ \cite{Azeyanagi:2007bj}. Since the information within the matter CFT moves on null rays, the information inside the interval can be mapped into the time interval $\left[t-z,t+z\right]$. One can view this as the boundary observer's ignorance to information prior and after this interval (Figure \ref{SinCFT} left).
\\~\\
\noindent In quantum gravity, we generically expect the entanglement entropy formula \eqref{CFTS} to be qualitatively influenced in two ways. Firstly, gravitons also contribute to the entanglement entropy and they should be taken into account. Secondly, a conceptually deeper question is how one defines the location of the entangling surface invariantly within quantum gravity. For JT gravity, bulk gravitons are of course absent, but we can deal with the second conceptual issue in a precise way. \\
The location of the entangling surface is at the bulk point $(u,v)$, which is found by the radar definition from the boundary observer's times $u$ and $v$. Given two boundary times $u$ and $v$, the entanglement entropy can be viewed as a diff-invariant bulk observable if we construct it as $S_{\text{ent}}[f(u),f(v)]$, and it is this operator that we will insert in the gravitational path integral \eqref{pischw}. 
The computation can be done by taking the $\ell$-derivative of the two-point function \eqref{sch2pt} and setting $\ell=0$ in the end:
\begin{equation}
\label{Squa}
\left\langle S_{\text{ent}}\right\rangle_{\beta} +\frac{c}{6}\ln \delta = -\frac{c}{12} \frac{1}{Z}\int d\mu(k_1) d\mu(k_2)  e^{i 2z \frac{k_1^2}{2C}}e^{-(\beta+i2z) \frac{k_2^2}{2C}}\left.\frac{\partial}{\partial \ell}\left(\frac{\Gamma(\ell \pm i k_1 \pm i k_2)}{(2C)^{2\ell}\Gamma(2\ell)}\right)\right|_{\ell=0},
\end{equation}
time-independent as it should be. For small separations $z \ll C$, we retrieve the semi-classical formula \eqref{Ssemi}. \\
Still for a macroscopic black hole but in the very-near horizon regime $\beta \ll C \ll z$, the $k_2$-integral is dominated by its saddle (leading to the equivalence between microcanonical and canonical ensembles), whereas the large $z$-regime enforces $k_1 \approx k_2$. Using an integral formula for the Gamma-functions, the computation is identical to that in \cite{zhenbin}\footnote{In fact, the computation of \cite{zhenbin} computes the same quantity but taking $t\to t + i\beta/2$ to reach the other side of the TFD. This computes the wormhole length and is interpreted as the computational complexity.} and, upon subtracting the UV-divergent piece, leads to 
\begin{equation}
\left\langle S_{\text{ent}}\right\rangle_{\beta} \sim \frac{c}{6}\frac{2\pi}{\beta} z,
\end{equation}
of the same form as \eqref{Sth}. Hence the linear increase of the entanglement entropy happens well past the semi-classical regime. This suggests the identification of the total thermal entropy in the thermal atmosphere and the entanglement entropy of the half-space remains true for macroscopic black holes $\beta \ll C$, in spite of probing the dangerous deep bulk $z \gg C$ where quantum gravitational effects are expected, see also \cite{Blommaert:2019hjr}.
\\~\\
From the perspective of the 0+1d boundary theory, the full entropy contains the classical Bekenstein-Hawking contribution, combined with the bulk entanglement entropy as in the FLM-framework \cite{Faulkner:2013ana,Jafferis:2015del}. Since in the JT-model, Newton's constant scales as $G_N \sim 1/C$, the contribution \eqref{Ssemi} with $z\to+\infty$ is $\sim G_N^0$, with its quantum corrections in \eqref{Squa} contained as a series expansion in $1/C$. From the holographic perspective, we have to introduce counterterms to remove the $z$-divergent terms. For the semi-classical piece \eqref{Ssemi}, we would end with only a logarithmic contribution $\sim \frac{c}{6} \ln T$, which was studied from a (non-)decoupling argument perspective in \cite{Spradlin:1999bn}, and thermodynamically in \cite{Almheiri:2014cka} from the holographic boundary stress tensor.  \\
Hence, the expression \eqref{Squa} contains a partial summation of all diagrams contributing to the full entropy, namely those incorporating matter-gravity interactions. The contribution that is left out is the pure gravity piece, coming solely from the boundary graviton degree of freedom, which is an edge state contribution, see e.g. \cite{Lin:2018xkj,Blommaert:2018iqz}. This pure gravity entropy is just the thermal entropy of the Schwarzian and is one-loop exact:
\begin{equation}
S_{\text{grav}} = \frac{4\pi^2 C}{\beta} + \frac{3}{2} + \frac{3}{2}\ln \frac{2\pi C}{\beta}.
\end{equation}
The first term is the classical Bekenstein-Hawking term, and the other terms can be treated on the same footing as the matter contributions \eqref{Ssemi}.\footnote{Recalling the fact that the matter action does not affect any pure gravity computation (see footnote \ref{fn1}), additional matter loop corrections to $S_{\text{grav}}$ seem to be absent.} Notice that the pure gravity sector also contributes a term $\sim \ln T$ as a one-loop contribution, similarly to the matter sector. We leave a more thorough study to future work.
\\~\\
For completeness, this set-up can be readily extended to the entanglement entropy for a bulk interval (Figure \ref{Sentbulk}).
One uses the following quantum gravity operator insertion:
\begin{equation}
S_{\text{ent}}[f(u_1),f(v_1),f(u_2),f(v_2)] = \frac{c}{12}\ln \frac{(f(u_1)-f(u_2))^2}{\delta^2f'(u_1)f'(u_2)} + (u \leftrightarrow v) + \frac{c}{6}\ln \eta + \ln G(\eta),
\end{equation}
where $\eta = \frac{(f(u_1)-f(v_1))(f(u_2)-f(v_2))}{(f(u_1)-f(v_2))(f(u_2)-f(v_2))}$ is the reparametrized cross-ratio, and $G(\eta)$ is a non-universal function that is only known in very specific cases, see \cite{Almheiri:2019psf} for a discussion in this context.\footnote{Considering only the first two terms of this expression, corresponds to the operator required when the boundary is transparent instead of reflecting. Semi-classically, we will use this operator in the next section. The quantum treatment would be harder though, since the action is not just the Schwarzian action.} As before, this quantity is independent of the precise spatial form of the Cauchy surface $\Sigma$ on which the entropy is computed. The resulting quantity is a bulk bilocal observable, whose Schwarzian path integral is more difficult to compute. It is amusing to note that the log of a crossratio can be computed explicitly using the techniques of \cite{Blommaert:2019hjr}.

\section{Semi-classical entanglement of Hawking particles}
\label{sect:evap}

\begin{figure}[h]
\begin{center}
\medskip
\includegraphics[width=0.8\textwidth]{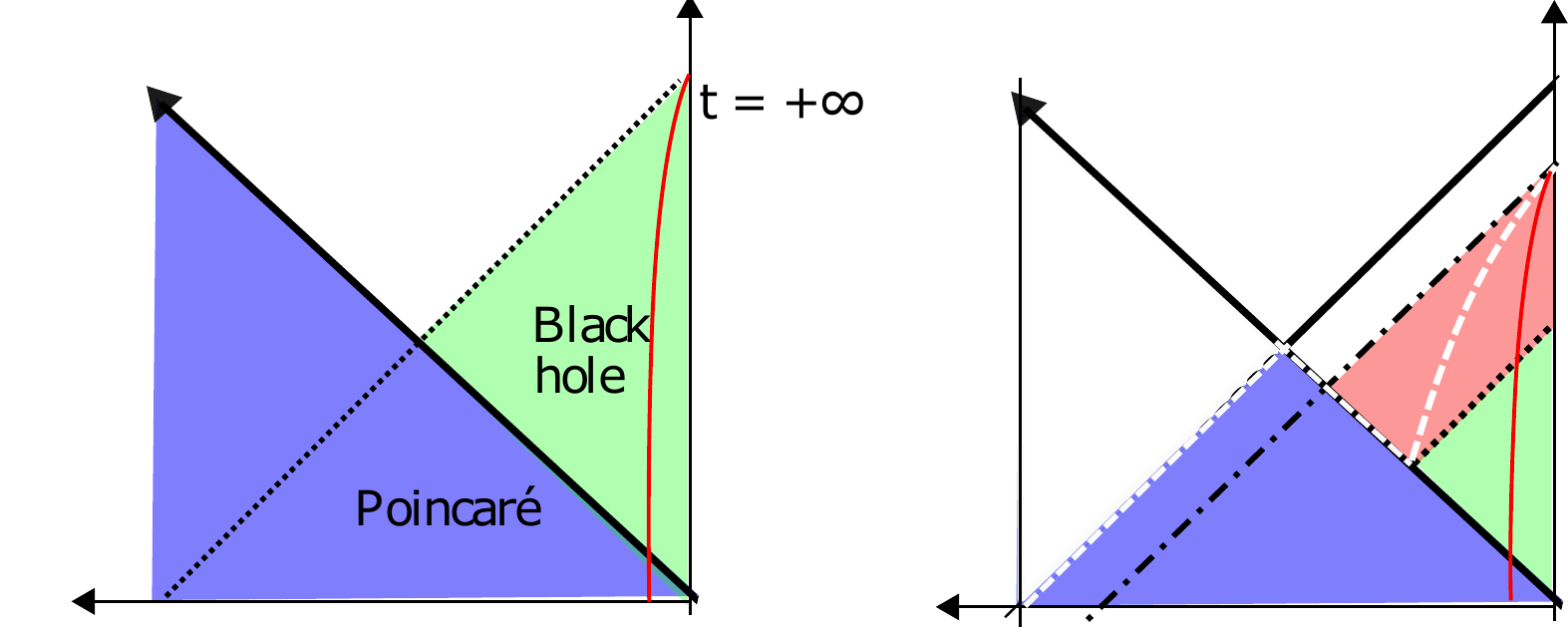}
\end{center}
\vspace{-2mm}
\caption{Left: Creation of a black hole by sending in a pulse in the Poincar\'e patch. The dashed line represents the black hole horizon as described in the black hole frame. The red line is the holographic boundary curve. Right: Evaporating black hole structure. The blue region is the original Poincar\'e patch. After the initial pulse passes, the boundary observer lives in the red evaporating patch which encompasses the green (non-evaporating) frame. The white dashed line represents the apparent horizon which jumps from the initial Poincar\'e extremal horizon to the would-be horizon in the non-evaporating case and then recedes back as the evaporation continues.}
\label{fig:blackhole}
\end{figure}

Everything up to this point concerned non-evaporating black holes, as black holes in AdS tend to equilibrate instead. In order to allow evaporation, we have to modify the asymptotic boundary conditions from perfect reflection to absorption. This model was studied in \cite{ads2}, and we retake it here.\\
Energy conservation dictates that the total bulk energy can only be modified by in- and outfluxes of matter in the sense that:
\begin{equation}
\label{encon}
\frac{dE}{dt} = \normal{T_{vv}(t)} \hspace{-0.1cm}- \hspace{-0.1cm} \normal{T_{uu}(t)}.
\end{equation}
Consider now an infalling matter pulse at $t=0$, with hence $T_{vv}(t) = E_0\delta (t)$. Solving the Schwarzian equation of motion, this causes us to transfer from the Poincar\'e solution $f(t)=t$ to the thermal solution $f(t) = \tanh \sqrt{\frac{E_0}{2C}} t$ (Figure \ref{fig:blackhole} left). 

\begin{figure}[H]
\centering
\includegraphics[width=0.55\textwidth]{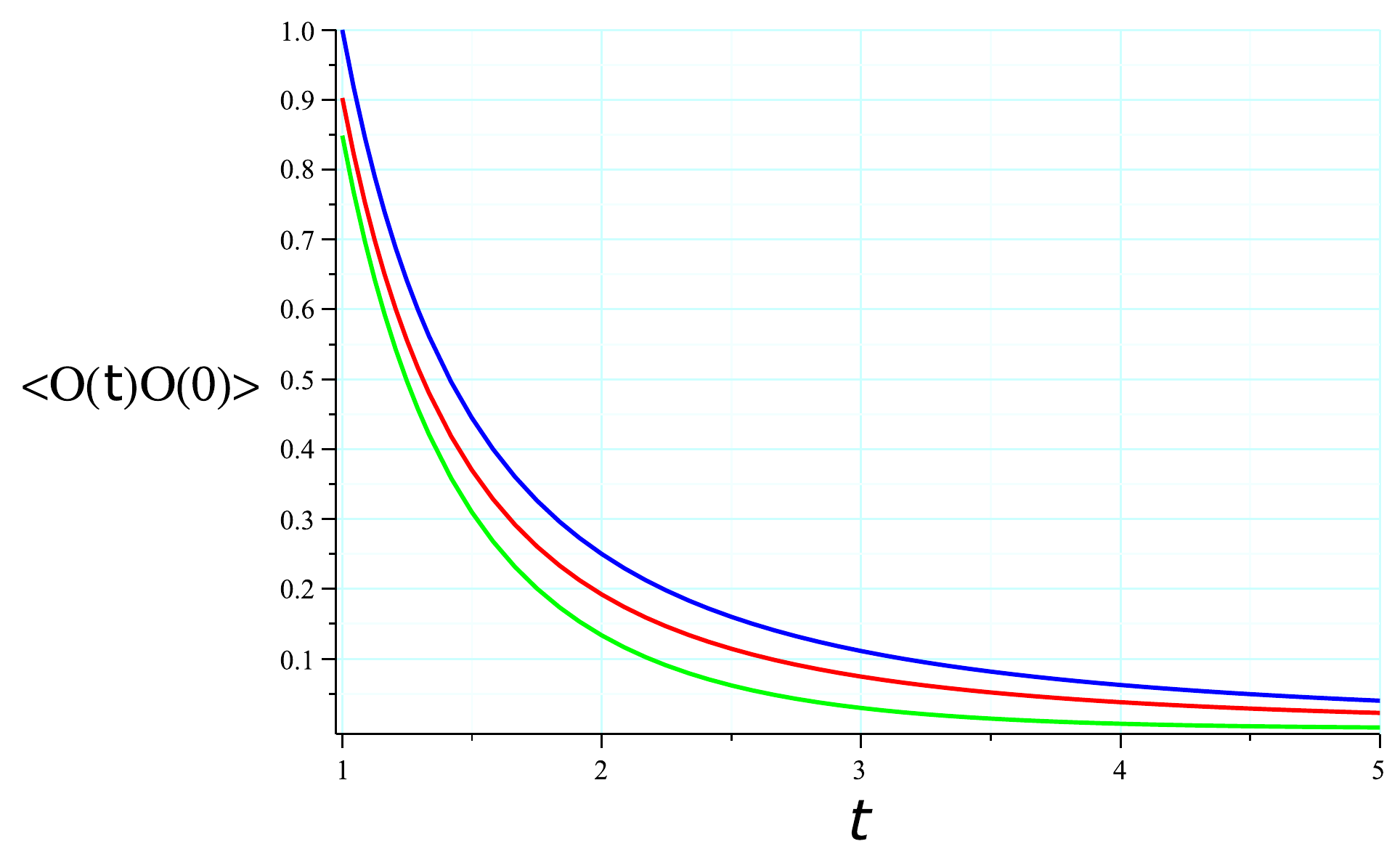}
\caption{Two-point correlator for a field with $\ell=1$ and $C=1/2$. Blue: Poincar\'e background $f(t)=t$. Green: Eternal black hole $f(t) = \tanh \sqrt{E_0}t$. Red: evaporating black hole $f(t) = \eqref{exsolu}$.}
\label{decay}
\end{figure}

\noindent Besides this pulse, we set the boundary conditions such that nothing reflects back into the bulk: $\normal{T_{vv}(t)}=0$ for $t>0$ as perfect absorption boundary conditions. Both the total boundary energy $T_{tt}(t) =  -C \left\{f,t\right\}$ and the ingoing flux $\normal{T_{uu}(t)} = -\frac{c}{24\pi}\left\{f,t\right\}$ are given by Schwarzian derivatives. Hence plugging these in \eqref{encon}, the energy decays exponentially in the system:
\begin{equation}
\label{massbh}
T_{tt}(t) = E_0 e^{-At},
\end{equation}
where $A=\frac{c}{24\pi C}$. This leads to the time reparametrization profile \cite{ads2}:\footnote{The expression in the second line was recently written down in \cite{Almheiri:2019psf}.}
\begin{align}
\label{exsolu}
f(t) &=  \frac{1}{\alpha^2} \int_0^t dx \frac{1}{\bigl(I_1\left(\alpha \right)K_0\bigl(\alpha e^{-\frac{Ax}{2}}\bigr) + K_1\left(\alpha \right) I_0\bigl(\alpha e^{-\frac{Ax}{2}}\bigr) \bigr)^2} \nonumber \\
&= \frac{2}{\alpha A} \frac{I_0(\alpha) K_0(\alpha e^{-At/2}) - K_0(\alpha)I_0(\alpha e^{-At/2})}{I_1(\alpha) K_0(\alpha e^{-At/2}) + K_1(\alpha)I_0(\alpha e^{-At/2}},
\end{align}
with $\alpha = \frac{24\pi}{c} \sqrt{2C E_0}$. This time reparametrization asymptotes to a fixed value beyond the eternal black hole horizon. The endpoint however does not reach the original Poincar\'e horizon. The Penrose diagram of the evaporating hole is shown in Figure \ref{fig:blackhole} right.
\\~\\
In any given frame $f$, the semi-classical two-point correlator is of the form:
\begin{equation}
\left\langle \mathcal{O}(t_1)\mathcal{O}(t_2)\right\rangle = \left(\frac{f_1' f_2'}{(f_1-f_2)^{2}}\right)^\ell.
\end{equation}
In the evaporating black hole frame \eqref{exsolu}, two-point functions decay at an intermediate pace between the polynomial decay in the vacuum and the exponential decay in the eternal black hole (Figure \ref{decay}). 

\begin{figure}[h]
\centering
\includegraphics[width=0.9\textwidth]{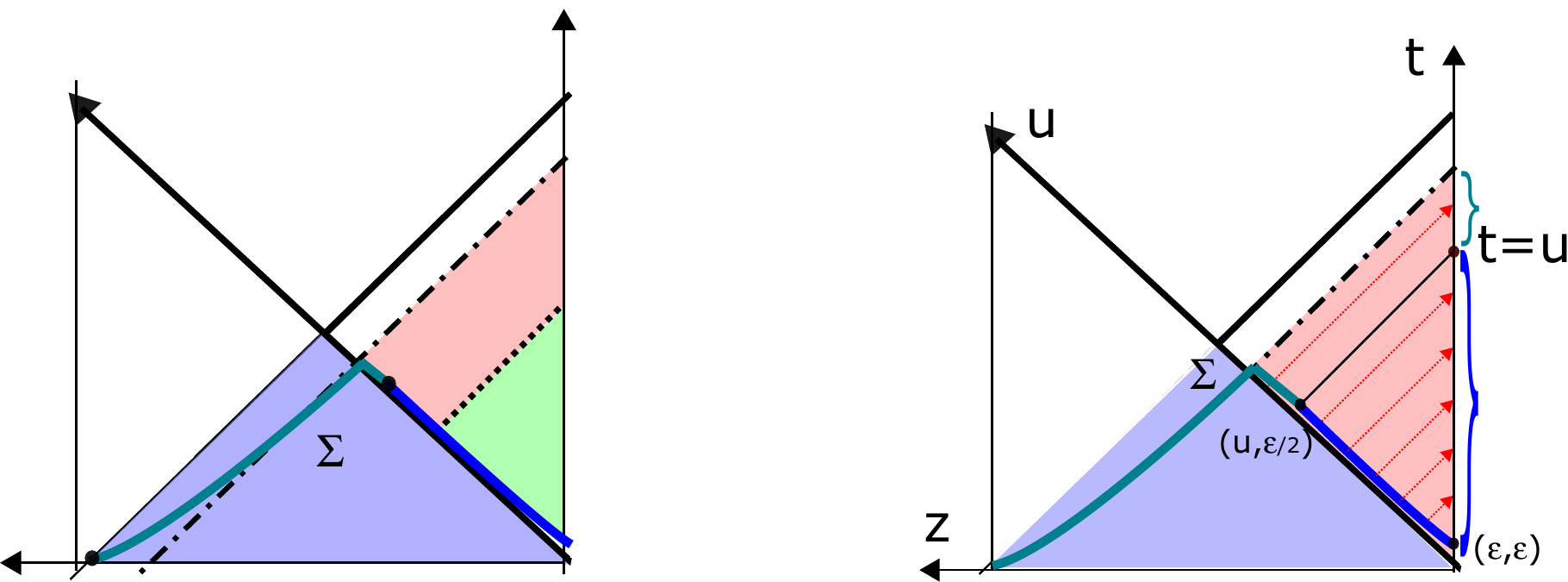}
\caption{Left: Evaporating black hole structure, with a spacelike Cauchy surface $\Sigma$ superimposed. The surface is almost lightlike, and is divided in two pieces (blue and darkgreen). One computes the entanglement entropy between the two pieces, as the division point moves closer and closer to the final event horizon as boundary time progresses. Right: Early-late entanglement of Hawking quanta at the time $t$ can be computed by computing the bulk entanglement entropy between both parts of the bulk Cauchy surface $\Sigma$.}
\label{singappv4}
\end{figure}

\begin{figure}[h]
\centering
\includegraphics[width=0.45\textwidth]{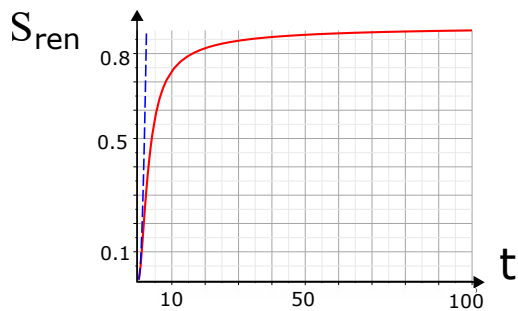}
\caption{Renormalized entanglement entropy \eqref{final} as a function of boundary time $t$ for $A=1$, $\alpha=\sqrt{2}$ and $c=12$. The dashed blue line represents the same quantity for the non-evaporating black hole, which after starting in the same way, diverges linearly for large $t$.}
\label{Sent}
\end{figure}

\begin{figure}[h]
\centering
\includegraphics[width=0.45\textwidth]{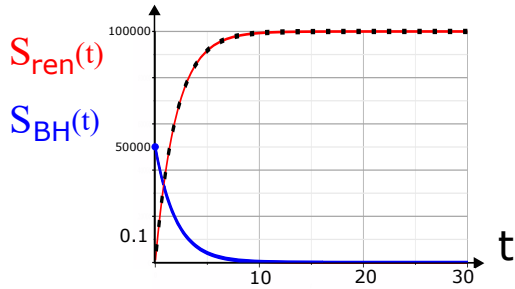}
\caption{Renormalized entanglement entropy \eqref{final} and black hole entropy \eqref{SBH} as a function of boundary time $t$ for $A=1$, $\alpha=50000$ and $c=12$. The dotted black line is the coarse-grained matter entropy, the r.h.s. of \eqref{Scg}.}
\label{SentCla}
\end{figure}

\noindent Let us now use this solution to compute the entanglement entropy in the matter sector between the early and late Hawking radiation. This computation is similar to that of \cite{Almheiri:2013wka} done for the asymptotically flat CGHS model. Where JT gravity stands out again, is in the full analytic solvability of the problem. \\
The entanglement entropy of the matter fields can be computed using a Cauchy surface $\Sigma$ that is close to the initial infalling pulse and then reconnects to the Poincar\'e horizon in the end (Figure \ref{singappv4}). Note that we are required to study a Cauchy surface in the entire Poincar\'e patch as this is our starting geometry.
\noindent The interval is between $(u=\epsilon, v=\epsilon)$ and $(u=t, v=\epsilon/2)$ for some infinitesimal $\epsilon$ whose sole purpose is to make sure the surface is spacelike.
As the matter is null and propagates unhindered to the timelike boundary, the entanglement entropy across this interval on $\Sigma$ measures the entanglement entropy the boundary observer would associate to all radiation he received until some specific time $t$, measured in his evaporating time frame (see also figure \ref{SinCFT} right). \\
The entanglement entropy in the interval in the Poincar\'e vacuum state, but described using the evaporating frame \eqref{exsolu} is then found as
\begin{equation}
S = \frac{c}{12} \ln \frac{(f(u_1)-f(u_2))^2}{\delta^2 f'(u_1)f'(u_2)} + \frac{c}{12} \ln \frac{(f(v_1)-f(v_2))^2}{\delta^2 f'(v_1)f'(v_2)}.
\end{equation}
The UV-cutoff $\delta$ is the one used by the time frame $t$ and hence the one of the local boundary observer. Since $u_2\approx v_2 \approx v_1 \approx 0$, plugging in the values $f(u_2)=f(v_2)=f(v_1)=0$, $f'(u_2)=f'(v_2)=f'(v_1)=1$ and subtracting the entanglement entropy in the evaporating state,\footnote{The boundary observer's detector is calibrated to the vacuum he would define using his time coordinate: the evaporating state. This is analogous to the flat space Unruh effect where one considers the Minkowski vacuum to find the thermal population in Rindler coordinates, upon subtracting the Rindler vacuum contribution. The Poincar\'e state plays the role of the Minkowski vacuum. }
\begin{equation}
S_{\text{ref}} = \frac{c}{12} \ln \frac{(u_1-u_2)^2}{\delta^2} + \frac{c}{12} \ln \frac{(v_1-v_2)^2}{\delta^2},
\end{equation}
we find:
\begin{equation}
\label{final}
S_{\text{ren}} \equiv S - S_{\text{ref}} = \frac{c}{12} \ln \frac{f(t)^2}{ t^2f'(t)}.
\end{equation}
Since the outgoing radiation propagates along a null line, this same entanglement entropy is found on the boundary line as that distinguishing the early radiation (times  $<t$) with the late radiation yet to come (times $>t$) (Figure \ref{singappv4} right).
\\~\\
Plotting this function explicitly, one finds the result of Figure \ref{Sent}.
The entanglement entropy increases monotonically as time progresses, and information does not come out.
The mass of the remaining black hole \eqref{massbh} becomes arbitrarily low, in spite of the entanglement entropy not decreasing. 
This arises fully from the fact that $f(t)$ does not asymptote to $t$ itself: in the distant future, the frame is not reduced to the Poincar\'e patch (Figure \ref{singappv4}). If it were, then $S_{\text{ren}}$ would vanish in the end as well, leading to the information being returned with the Hawking radiation. \\
Hence we do not obtain a Page curve, and information is lost. A similar feature was observed with the analysis of the CGHS model \cite{Almheiri:2013wka}. In both cases, this can be viewed as a quantitative confirmation that the semi-classical Hawking computation is not able to restore information, but this is an artifact of the semi-classical approximation, see also \cite{Anous:2016kss}.
\\~\\
As a comparison, doing the same computation for the non-evaporating black hole, by plugging $f(t)=\tanh \sqrt{\frac{E_0}{2C}} t$ into \eqref{final}, one finds an ever-increasing entropy:\footnote{This expression is formally identical to that of the previous section upon setting $t \to 2z$ and $\sqrt{\frac{E_0}{2C}} \to \frac{\pi}{\beta}$. The volume-scaling $\sim z$ of the thermal matter entropy explains the linear time-scaling $\sim t$ found here.}
\begin{equation}
S_{\text{ren}} = \frac{c}{6}\ln \left(\sqrt{\frac{2C}{E_0}}\frac{1}{t}\sinh \left(\sqrt{\frac{E_0}{2C}}t\right)\right),
\end{equation} 
for large $t$ scaling linearly in time: $S_{\text{ren}} \sim t$. The perfectly thermal Hawking emission does not contain any information whatsoever, and the information contained in the compensating ingoing quanta is lost in the process, see also \cite{Fiola:1994ir}. This quantity is also plotted in Figure \ref{Sent}.
\\~\\
One checks explicitly using \eqref{exsolu} that $f(+\infty) = \frac{2}{\alpha A}\frac{I_0(\alpha)}{I_1(\alpha)}$ is finite and $f'(+\infty) \sim 1/t^2$, leading to a constant value of $S_{\text{ren}} \to \frac{c}{6}\ln I_0\big(\frac{24\pi}{c}\sqrt{2CE_0}\big)$ at late times, identifiable with the amount of lost information in the evaporation processs. In particular, for a macroscopic black hole where $E_0 \gg 1/C$, and hence $\alpha \gg 1$, one can approximate the late-time value as
\begin{equation}
S_{\text{ren}}(t \to \infty)\, \to \, 4 \pi \sqrt{2C E_0} = 2 S_{\text{BH}}(t=0),
\end{equation}
which is \emph{twice} the original Bekenstein-Hawking entropy of the formed black hole. This factor of two has been found before in the similar context of the CGHS / RST model \cite{Fiola:1994ir}, and the physical interpretation is the following. It was shown by Zurek a long time ago \cite{Zurek:1982zz}, that a black hole evaporating into empty space is an irreversible process where the final thermodynamic (coarse-grained) matter entropy is larger than the initial black hole entropy, by a factor of $(D+1)/D$ in $D$ spatial dimensions. This gives a factor of two for $D=1$. The argument basically compares the infinitesimal decrease in BH entropy: $\delta S_{\text{BH}} = -\frac{\delta E}{T_H} $, with the increase in thermal entropy of a free Bose gas in $D$ dimensions in a time span $dt$ in the heat bath generated by the black hole itself: $\delta S = \frac{D+1}{D} \frac{E}{T_H} dt$ with $\delta E = E dt$.\footnote{Since the (semi-classical) Planck black body law still holds for the massless scalar in AdS$_2$, this argument is unchanged from the flat case. Moreover, no greybody factors appear in 2d.} \footnote{As a sidenote, if one would demand this evaporation to happen \emph{reversibly} to interpret the black hole microstates, one would need to take a matter system satisfying the thermodynamic relation $S = \frac{E}{T}$. An example is given in \cite{Zurek:1982zz} by placing the black hole in an external bath at temperature very near $T_H$. Even stronger, if one would demand the process of emitting a small pocket of radiation to relate matter equilibrium configurations, one needs the more stringent $S = \frac{E}{T_H}$, which is a Hagedorn system of long strings with $T_{\text{Hag}} = T_H$ \cite{Mertens:2015hia}.}\\
Here, however, the factor of two appears for the fine-grained (entanglement) entropy. One can physically interpret this as the thermal entropy of the emitted gas being dominated by the cross-horizon correlations, leading to an equality between fine- and coarse grained matter entropies. \\
In fact, still for a macroscopic black hole for which $E_0 \gg 1/C$, the equality of these entropies holds for all times. The thermal entropy of the black hole as it evaporates is given by:
\begin{equation}
\label{SBH}
S_{\text{BH}}(t) = 2\pi \sqrt{2C E_0}e^{-\frac{At}{2}}.
\end{equation}
Using the asymptotic forms of the Bessel functions in \eqref{exsolu} for $\alpha \gg 1$, one can approximate the fine-grained entropy as:
\begin{equation}
\label{Scg}
S_{\text{ren}}(t) \approx 4\pi \sqrt{2CE_0}\left( 1- e^{-\frac{At}{2}}\right) = 4\pi \sqrt{2CE_0} - 2 S_{\text{BH}}(t),
\end{equation}
indeed satisfying Zurek's irreversibility argument $\delta S_{\text{ren}} = -2\delta S_{\text{BH}}$ for the thermal entropy at all times. We plotted these different entropy functions in Figure \ref{SentCla}.

\noindent Defining the Page time as the time when the thermal entropies of the black hole and the radiation are the same, we obtain the Page time
\begin{equation}
t_{\text{Page}} = \frac{48 \pi }{c} C \ln \frac{3}{2}.
\end{equation}

\noindent We have seen in the previous sections how Schwarzian techniques can be used to go beyond semi-classical gravity. What is required here, is an embedding of this computation within a unitary quantum-mechanical framework. Due to the absorbing boundary conditions however, this is not so simple, and is postponed to future work.

\section{Concluding remarks}
The Unruh and Hawking effects are of fundamental importance in understanding quantum black holes, but unfortunately it is very difficult to go beyond the level of matter quantum fields in a curved spacetime. Due to its solvability, Jackiw-Teitelboim gravity is an ideal test-case to attempt to include quantum gravitational effects, which we have studied throughout this work. 
\\~\\
Within the set-up of a thermal quantum system, we have studied several bulk diff-invariant operators: the bulk stress tensor components $:T_{uu}[f(u)]:$, $:T_{vv}[f(v)]:$, the spectral occupation number $N_\omega[f]$ and the matter entanglement entropy $S_{\text{ent}}[f(u),f(v)]$. We defined these objects operationally using only boundary-intrinsic data and studied how quantum gravitational effects modify them from their semi-classical limit. It would be interesting to study more general correlation functions where several of these objects, combined with boundary bilocals $\mathcal{O}_\ell (t_1,t_2)$ and local HKLL bulk fields \cite{Blommaert:2019hjr}, are applied together.
\\~\\
In the final section, we imposed absorbing boundary conditions at the holographic boundary to allow the bulk black hole to evaporate. We computed the entanglement between the early and late Hawking radiation in this model and found information loss within the semi-classical set-up. This was expected and illustrates that unitarity can seemingly be violated as an artifact of the semi-classical perturbative expansion, see however \cite{Almheiri:2019psf,Penington:2019npb}. It would be very interesting to combine these two types of calculations, and reach a quantum understanding of evaporation in this model. This is left to future work.
\\~\\
The 1+1d JT model is actually very universal in several ways. Firstly, it describes the low-energy dynamics of SYK-like systems. Secondly, it corresponds to the s-wave sector of pure 3d $\Lambda < 0$ gravity (see e.g. \cite{Achucarro:1993fd} for the original argument, and \cite{Mertens:2018fds} for a discussion in this context). Finally, near-extremal (charged and/or rotating) black holes in higher dimensions develop a long throat with an AdS$_2$ near-horizon region (in product with some compact space), whose dynamics is governed by JT gravity. Within this set-up, it has been argued that one can think of the Schwarzian wiggly boundary curve as separating the near-horizon JT region from the asymptotic region \cite{Nayak:2018qej,Sachdev:2019bjn,Moitra:2018jqs,Moitra:2019bub}. As such, it is intriguing to contemplate having defined the near-horizon frame and (s-wave) observables (studied in this paper) for a higher-dimensional black hole using this separating curve as an anchor. \\
The analysis of the last section describes only entanglement of s-wave Hawking particles of higher-dimensional black holes. We generically expect particles with other angular momenta to experience a gravitational potential hindering their escape. They are moreover more energetic and hence Boltzmann-suppressed. Hence the restriction to the s-sector seems plausible to obtain the essential horizon entanglement physics. \\
These links with other systems illustrate the potential applications of any study done within JT gravity, either for higher-dimensional black holes or for full-fledged holographic systems (e.g. SYK). It would be interesting to understand this better.

\section*{Acknowledgements}
%%%%%%%%%%%%%%%%%%%%%%%%%%%%%%%%%%%%%%%%%%%%
I thank A. Blommaert, N. Callebaut, J. Engelsoy, A. Kamenev, P. Nayak and H. Verlinde for discussions. The author gratefully acknowledges financial support from Research Foundation Flanders (FWO Vlaanderen).

\appendix
\section{Time-dependent Schwarzian coupling and Ward identities}
\label{app:timed}
We study Schwarzian QM with a classical time-varying coupling constant $C(\tau)$, with action
\begin{equation}
\label{Schwgen}
S = -\int d\tau \, C(\tau) \left\{f,\tau\right\},
\end{equation}
and use this to derive the Ward identities for stress tensor insertions in correlation functions. We work in Euclidean signature in this appendix. 

\subsection{Time-dependent couplings and einbeins}
The Hamiltonian corresponding to \eqref{Schwgen} is time-dependent. Defining a new time variable $\tilde{\tau}$ as
\begin{equation}
\label{timerep}
d\tilde{\tau} = \frac{d\tau}{2C(\tau)} \quad \Rightarrow \quad \tilde{\tau} = \int^{\tau} \frac{d\tau}{2C(\tau)},
\end{equation}
the generalized Schwarzian action \eqref{Schwgen} transforms into a regular one:
\begin{equation}
S = -\frac{1}{2}\int d\tilde{\tau} \, \left\{f,\tilde{\tau}\right\} - \frac{1}{2}\int \frac{C'(\tau)^2}{C(\tau)}d\tau,
\end{equation}
which we know how to compute with.\footnote{The second term does not depend on the reparametrization $f$ and only gives an overall factor which drops out when we normalize correlators with the vacuum amplitude. It is not interesting for our purposes. I thank A. Kamenev for a discussion on this formula.} As well-known, the Schwarzian model has no time reparametrization invariance (1d diff invariance).\footnote{From a 2d CFT perspective, conformal invariance is completely broken explicitly when doing the dimensional reduction, and only the Virasoro zero-mode $L_0$ survives.} Bilocal operators are transformed by \eqref{timerep} as
\begin{equation}
\left(\frac{\partial_\tau{f}_1\partial_\tau{f}_2}{(f_1-f_2)^2}\right)^{\ell} = \frac{1}{(2C(\tau_1))^\ell (2C(\tau_2))^\ell}\left(\frac{\partial_{\tilde{\tau}}f_1 \partial_{\tilde{\tau}}f_2}{(f_1-f_2)^2}\right)^{\ell}.
\end{equation}
The resulting vacuum two-point function for a given profile $C(\tau)$ then becomes:
\begin{equation}
\label{zerocor}
G^\infty_{\ell,C(\tau)}(\tau_1,\tau_2) = \int dk^2\sinh{(2 \pi  k)}\, e^{ - (\tilde{\tau}(\tau_2)-\tilde{\tau}(\tau_1)) k^2}\, \frac{ \Gamma^2( \ell+ i k)\Gamma^2(\ell- i k)}{(2C(\tau_1))^\ell (2C(\tau_2))^\ell \,2\pi^2\, \Gamma(2\ell)},
\end{equation}
with $\tilde{\tau}(\tau)$ as given in \eqref{timerep}. This generalizes immediately to higher-point functions. Obviously, time-translation invariance $\tau \to \tau +c$ is lost for any non-constant profile $C(\tau)$. 
This simple generalized model also demonstrates that one can compute amplitudes using quasi-statics of the coupling $C(\tau)$: no dependence on its derivatives is present. This resonates with the exactness of the quasi-static approximation in 2d JT gravity, as we mentioned in Sections \ref{sec:pulse} and \ref{sect:evap}.
\\~\\
This procedure also has an interpretation in JT gravity \cite{ads2,Almheiri:2014cka,Jensen:2016pah,Maldacena:2016upp}. Changing the asymptotic boundary value of the dilaton as
\begin{equation}
\Phi = \frac{a}{\tilde{z}} \quad \to \, \frac{a(\tau)}{z},
\end{equation}
with $z=\epsilon \dot{\tau}$ and $\tilde{z}=\epsilon \dot{\tilde{\tau}}$, requires the time reparametrization
\begin{equation}
\frac{d\tau}{d\tilde{\tau}} = a(\tau)/a,
\end{equation}
and indeed transforms the original Schwarzian action into the generalized one \eqref{Schwgen}, identifying $a(\tau)/a = C(\tau)$. Any choice of the dilaton asymptotics breaks explicitly 1d conformal invariance, where the different possible theories are parametrized by the given function $a(\tau)$. 

\subsection{Comments on Liouville embedding}
\begin{figure}[h]
\centering
\includegraphics[width=0.4\textwidth]{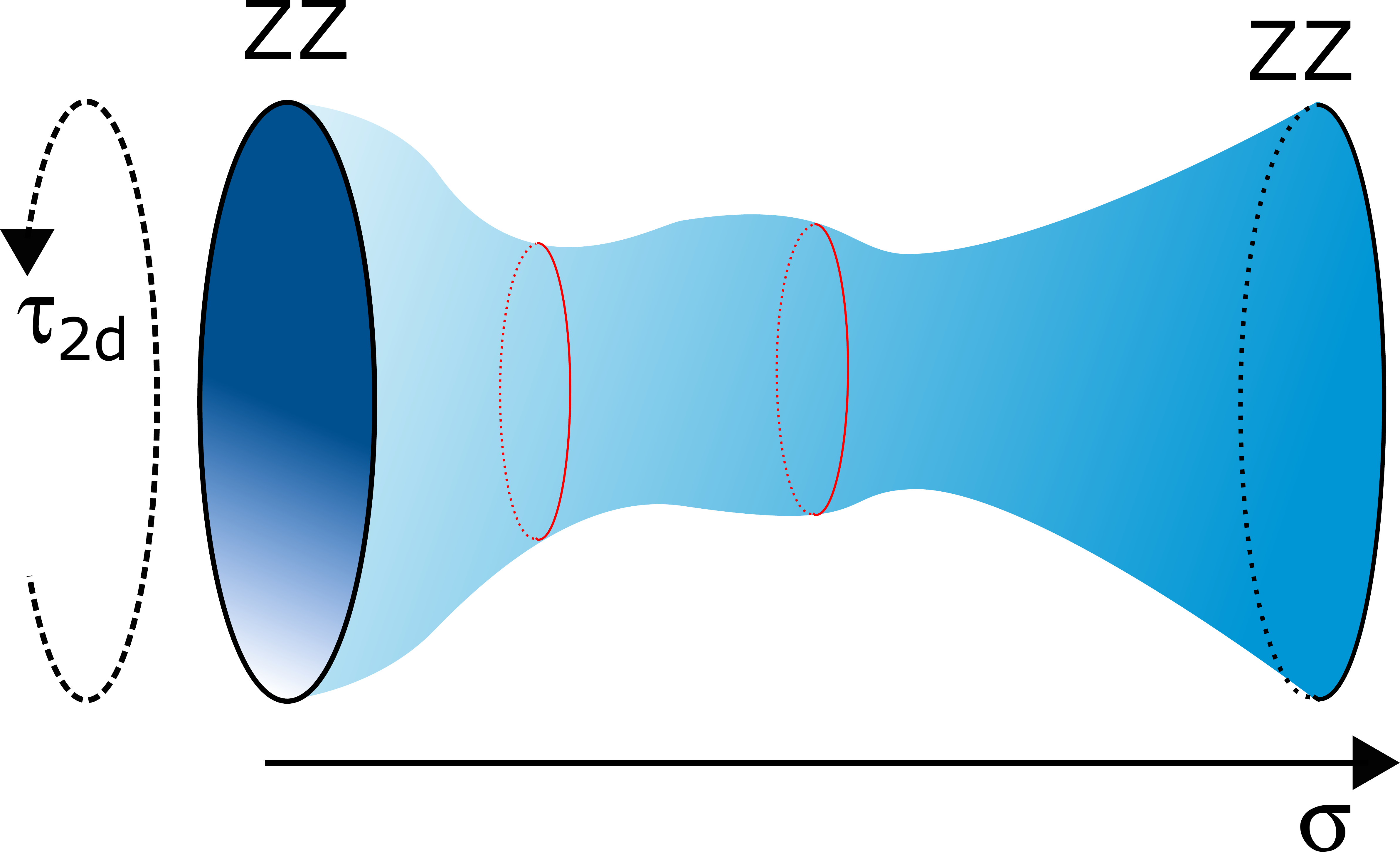}
\caption{Liouville ZZ-brane system with a varying circular circumference $T(\sigma)$ for the angular coordinate $\tau_{2d} \sim \tau_{2d} + T(\sigma)$.}
\label{ZZwiggle}
\end{figure}

When embedding the generalized Schwarzian theory \eqref{Schwgen} within Liouville theory \cite{Mertens:2017mtv,Mertens:2018fds} between two ZZ-branes, we should consider a cylindrical surface with a varying cylinder radius $T(\sigma)$, in the double-scaling limit $c\to+\infty$, $T(\sigma) \to 0$ keeping fixed $cT(\sigma) = C(\sigma)$ to some prescribed function (Figure \ref{ZZwiggle}). 
Within 2d CFT, a varying cylinder radius is trivial as it can be undone by a conformal transformation. However, when taking the Schwarzian limit, 2d conformal symmetry is taken to 1d reparametrization symmetry, which is explicitly broken in this procedure. \\
As emphasized in \cite{Mertens:2017mtv}, the Liouville computation reduces to a minisuperspace Hamiltonian propagation amplitude along the cylinder, where the Schwarzian time coordinate $\tau$ is identified with the Liouville $\sigma$-coordinate, $\tau \equiv \sigma$, and one finds a time-dependent Schwarzian coupling constant $C(\tau)$ in \eqref{Schwgen}. The changing circumference corresponds to a time-dependent Hamiltonian. The boundary ZZ-states and Liouville primary vertex operator insertions happen at a single instant in time $\tau$, and are unaffected by this varying radius. The only effect in the minisuperspace Liouville computation is then the replacement:
\begin{equation}
e^{-\tau H} \quad \to \quad  e^{- \int_0^{\tau}d\tau H(\tau)},
\end{equation}
indeed what was found in \eqref{zerocor}.

\subsection{Ward identities}
\label{app:ward}
We can use the time-dependent coupling $C(\tau)$ to derive the Ward identity for stress tensor insertions in bilocal correlation functions. A version of the Ward identity for solely stress tensor correlators was explored in \cite{Stanford:2017thb}. The Ward-like identities for bilocal correlators were derived in \cite{Mertens:2017mtv} using the embedding within Liouville CFT. Finally, another approach is being developed in \cite{nsv}.
\\~\\
Writing $C(\tau) \equiv C/e(\tau)$, with a dimensionless einbein $e(\tau) \equiv \sqrt{g_{\tau\tau}}$, we can write the Euclidean stress tensor $T^{\tau\tau} \equiv \frac{2}{\sqrt{g}} \frac{\delta S}{\delta g_{\tau\tau}}$ as
\begin{equation}
\label{stressbdy}
T_{\tau\tau} = e(\tau)^2 \frac{\delta S}{\delta e(\tau)} = C \left\{f,\tau\right\}.
\end{equation}
Differentiating \eqref{zerocor} w.r.t. $e(\tau)$ and setting $e(\tau) = 1$ at the end, we find:
\begin{align}
\left\langle \mathcal{T}\, T_{\tau\tau}(\tau) \mathcal{O}_\ell(\tau_1,\tau_2)\right\rangle &= \Big(-\frac{\ell}{C}\delta(\tau-\tau_1) - \frac{\ell}{C} \delta(\tau -\tau_2) - \theta(\tau_1< \tau <\tau_2)\partial_{\tau_{12}}\Big) G^\infty_{\ell,C}(\tau_1,\tau_2).
\end{align}
In Fourier space, the stress tensor insertion leads to an energy $\frac{k^2}{2C}$ in between the ends of the bilocal. This is the zero-temperature Ward identity.
\\~\\
At finite temperature, the stress tensor insertion is $T_{\tau\tau}(\tau) =  C \left\{\tan\frac{\pi}{\beta}f,\tau\right\}$. The time-dependent thermal generalization of \eqref{zerocor} is given by the expression:
\begin{equation}
\small
G^\beta_{\ell,C(\tau)}(\tau_1,\tau_2) = \int d\mu(k_1) \int d\mu(k_2)\, e^{ - \int_{\tau_1}^{\tau_2} \frac{d\tau}{2C(\tau)}(k_1^2-k_2^2)}\,  e^{-\int_{0}^{\beta} \frac{d\tau}{2C(\tau)} k_2^2}\frac{ \Gamma( \ell \pm i k_1 \pm i k_2)}{(2C(\tau_1))^\ell (2C(\tau_2))^\ell \, 2\pi^2 \, \Gamma(2\ell)},
\end{equation}
leading to the thermal Ward identity
\begin{align}
&\left\langle \mathcal{T}\, T_{\tau\tau}(\tau) \mathcal{O}_\ell(\tau_1,\tau_2)\right\rangle_\beta \nonumber \\
&= \Big(-\frac{\ell}{C}\delta(\tau-\tau_1) - \frac{\ell}{C} \delta(\tau - \tau_2) - \theta(\tau_1< \tau <\tau_2)\partial_{\tau_{12}} - \partial_\beta \Big) G^\beta_{\ell,C}(\tau_1,\tau_2),
\end{align}
in Fourier space leading to an energy $\frac{k_1^2}{2C}$ between the legs of the bilocal, and $\frac{k_2^2}{2C}$ outside.
\\~\\
Inverse Laplace transforming this expression, one finds the Ward identity in a fixed energy eigenstate $\left|k\right\rangle$ with energy $k^2/2C$, for which the last term can be rewritten as $\left\langle T_{\tau\tau}(\tau) \right\rangle_k\left\langle\mathcal{O}_\ell(\tau_1,\tau_2) \right\rangle_k$, in agreement with the result of \cite{Mertens:2017mtv}.

\section{Planckian spectrum for a charged system}
\label{app:charge}
\begin{figure}[h]
\centering
\includegraphics[width=0.2\textwidth]{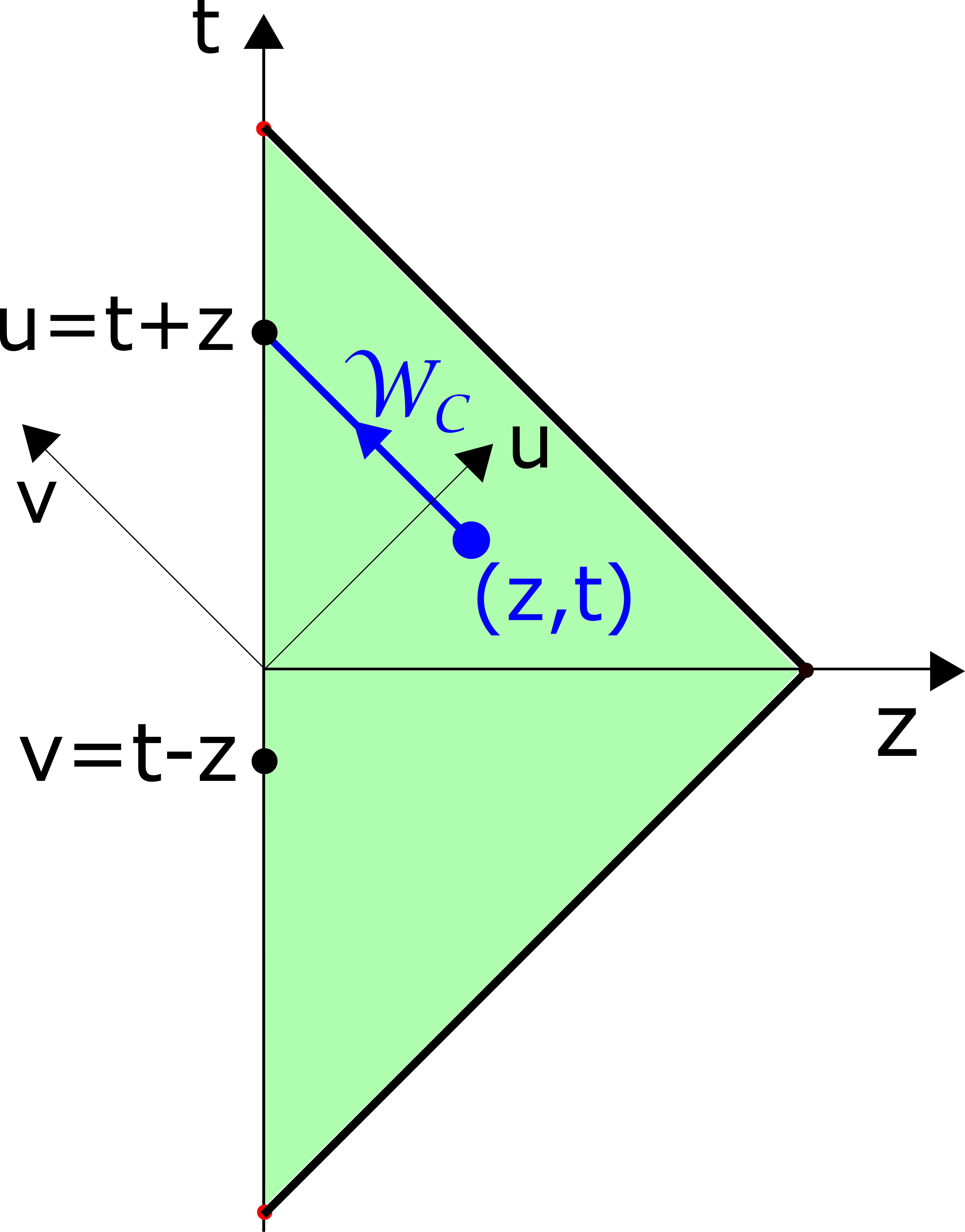}
\caption{Wilson line dressing to define a small-gauge-invariant bulk operator $\mathcal{O}_u (u,v)$. The right-moving operator $\mathcal{O}_v(u,v)$ would be constructed using a Wilson line along the $u$-direction, and leads to a factor $e^{i q \Lambda(v)}$ instead.}
\label{WilsonDress}
\end{figure}

Consider a bulk massless complex scalar field $\phi$, with action $S = \int d^2x \sqrt{-g} \partial_\mu \phi \partial^\mu \bar{\phi}$. Let $\phi$ be a charge $+q$ field, transforming as $\phi \to e^{iq \Lambda} \phi$. The grand canonical partition function $\mathcal{Z}(\beta,\mu) \equiv \text{Tr}\left[e^{-\beta H}e^{-\mu \beta Q}\right]$ of the matter sector can then be computed as the vacuum amplitude on the thermal manifold $\tau \equiv it \sim i t + \beta$ with twisted boundary conditions $\phi(\tau+\beta) = e^{- q \mu \beta} \phi(\tau)$ and $\bar{\phi}(\tau+\beta) = e^{+ q \mu \beta} \bar{\phi}(\tau)$. \\
Redefining the field using $\partial_\tau \phi = e^{- q \mu \tau} \partial_\tau \phi$ leaves the action invariant, and untwists the fields. The same is true when redefining the field using either the $u$ or $v$ light-cone coordinate. \\
This means the real-time two-point function in the grand canonical ensemble is readily computed as:
\begin{equation}
\label{chareclas}
\left\langle \partial_u \phi \, \partial_u \bar{\phi}\right\rangle_{\beta,\mu} = e^{i q \mu (u_1 - u_2)} \frac{1}{\frac{\beta^2}{\pi^2}\sinh\frac{\pi}{\beta}(u_1 - u_2)^2}.
\end{equation}
Semi-classically, the only difference is then the additional factor of $e^{i q \mu t}$ in \eqref{fouri}, leading to the shift $\omega \to \omega - q \mu$ in the Planckian spectrum \eqref{semiplanck}:
\begin{equation}
N_\omega[f(y)=y] \,=\,  \frac{\omega-q\mu}{\omega} \frac{e^{-\frac{\beta}{2}(\omega-q\mu)}}{e^{\frac{\beta}{2}(\omega-q\mu)}-e^{-\frac{\beta}{2}(\omega-q\mu)}},
\end{equation}
interpretable as the emission spectrum of a particle of energy $\omega$ and charge $q$ from a thermal system with temperature $\beta^{-1}$ and chemical potential $\mu$. Just as the bulk frame is fixed as the black hole frame $f(y)=y$, so also has the U(1) frame $\Lambda = 0$ been chosen, where $\Lambda$ is the gauge parameter: $A_\mu \to A_\mu + \partial_\mu \Lambda$. These frames are the classical solutions to the Schwarzian + free boson system.
\\~\\
Beyond classical gravity, we have to perform a path integral over frames. Reparametrizing to the diff-frame $f$ in \eqref{chareclas}, one writes
\begin{align}
\label{planckchem}
N_{\omega,q}[f] = -\frac{1}{\pi}\int dy_1 &\int dy_2 u_\omega(y_1) u^*_\omega(y_2) \left[e^{i q \mu (f(y_1) - f(y_2))}\frac{f'(y_1)f'(y_2)}{\frac{\beta^2}{\pi^2}\sinh^2 \frac{\pi}{\beta}(f_1-f_2)} - \left(\frac{1}{y_{12}}\right)^2\right].
\end{align}
However, this is not the end of the story since the bulk field $\phi[f](x)$ defined using the radar definition of the main text is still not observable as it carries charge. Including the coupling to the background gauge field, the bulk matter action is modified to $S = \int d^2x \sqrt{-g} D_\mu \phi \overline{D^\mu \phi}$ in terms of the gauge-covariant derivative $D_\mu = \partial_\mu - i q A_\mu$. In addition, the gauge theory itself is described by a BF-model, as relevant for the complex SYK-model \cite{Mertens:2018fds,Gaikwad:2018dfc,Sachdev:2019bjn,Moitra:2018jqs}, whose dynamics dictate that $F=0$ and the gauge field is hence pure gauge. The boundary dual is described by a free U(1) boson $\Lambda(t)$, coupled by the chemical potential $\mu$ to the gravitational degree of freedom $f(t)$. \\
To obtain an observable, several approaches can be followed. The simplest procedure is to extract the U(1) gauge-dependence of any covariant operator and fixing this to a predefined choice. E.g. the U(1) covariantly transforming $D_u \phi$ is taken to:
\begin{equation}
\label{chargedress}
D_u \phi \,\, \to \,\, e^{i q \Lambda(u)} \partial_u \phi,
\end{equation}
where the explicit gauge-dependence is fixed and extracted, and the remainder is in a gauge-fixed form. Here, due to $A$ being pure gauge in BF-theory, we can gauge-fix to $A=0$ in the last piece. \\
An equivalent prescription is to dress the operator with an EM Wilson line emanating from the boundary. A small-gauge-invariant operator is then constructed as
\begin{equation}
\label{chargedress}
\mathcal{O}_u (u,v) \equiv \mathcal{W}_\mathcal{C}D_u \phi = e^{i q \int^{u}_{v} dv A_v} D_u \phi =  e^{i q \Lambda(u)} \partial_u \phi,
\end{equation}
where we choose the bulk Wilson line to lie along a lightlike direction, connecting the bulk point $(u,v)$ with the boundary point $u$ or $v$, choosing the null direction opposite to the index in the operator (Figure \ref{WilsonDress}).\footnote{The middle equation shows the small-gauge-invariance of the operator. The r.h.s. uses the fact that $A$ is pure gauge in the bulk, allowing us to do a small gauge transformation to turn off $A$ at the bulk point, without changing the value of the operator. 
%\\
%As a longer alternative, we could extrapolate the boundary would-be gauge choice into the bulk by setting $A_v = \partial_v \Lambda(v)$ and $A_u = \partial_u \Lambda(v) = 0$, which indeed reduces at the boundary to $A_t = \partial_t \Lambda(t)$. With this (small) gauge choice, the equations of motion $D_uD_v \phi = \partial_u(\partial_v - iq A_v)\phi = 0$ are solved by the mode expansions:
%\begin{equation}
%\phi(u,v) = \sum_\omega a_\omega^- \frac{e^{-i\omega v}}{\sqrt{4\pi \omega}} + e^{iq \int^v dv A_v}\sum_\omega a_\omega^+ \frac{e^{-i\omega u}}{\sqrt{4\pi \omega}} + (a \leftrightarrow b^\dagger)
%\end{equation}
%One finds in particular $D_u \phi = e^{iq \Lambda(v)}\partial_u \phi^{A=0}$, in terms of the field $\phi^{A=0}$ without $A$-field turned on. Plugging this back into \eqref{chargedress}, one finds the r.h.s.
 %to set $A=0$ at the bulk point. Such a gauge choice keeps the distinction between left- and right-movers intact, and is very convenient in our case. Note that due to $F=0$, the Wilson line can be deformed keeping the endpoints fixed.
}
This leads to the (observable) two-point function in the generic diff-frame $f$ and U(1)-frame $\Lambda$:
\begin{equation}
\left\langle\mathcal{O}_u  \bar{\mathcal{O}}_u\right\rangle_{\beta,\mu}^{\text{CFT}} = e^{i q \mu (f(y_1) - f(y_2))}e^{i q (\Lambda(y_1) - \Lambda(y_2))}\frac{f'(y_1)f'(y_2)}{\frac{\beta^2}{\pi^2}\sinh^2 \frac{\pi}{\beta}(f_1-f_2)}.
\end{equation}
Alternatively, one can find both modifications simultaneously by using the boundary value $A_t = \mu \partial_t f + \partial_t \Lambda$ and extending this into the bulk. \\
By \eqref{chargedress}, the new observable operator $\mathcal{O}_u$ has a mode expansion related to the undressed operator $\partial_u \phi$ obtained by Fourier expanding $e^{iq \Lambda(u)}$, schematically:
\begin{equation}
\partial_u \phi = \sum_\omega a_\omega \partial_u u_\omega(u) + (a \leftrightarrow b^\dagger), \qquad \to \qquad \mathcal{O}_u = \sum_\omega \tilde{a}_\omega \partial_u u_\omega(u) + (\tilde{a} \leftrightarrow \tilde{b}^\dagger),
\end{equation}
in terms of new oscillators $\tilde{a}_\omega$. The result \eqref{planckchem} is now readily modified into:
\begin{align}
N_{\omega,q}[f,\Lambda] \equiv &\left\langle 0_F\right|\tilde{a}_\omega^{\dagger}\tilde{a}_\omega \left|0_F\right\rangle = -\frac{1}{\pi}\int dy_1 \int dy_2 u_\omega(y_1) u^*_\omega(y_2) \nonumber \\
&\times \left[e^{i q \mu (f(y_1) - f(y_2))}e^{i q (\Lambda(y_1) - \Lambda(y_2))}\frac{f'(y_1)f'(y_2)}{\frac{\beta^2}{\pi^2}\sinh^2 \frac{\pi}{\beta}(f_1-f_2)} - \left(\frac{1}{y_{12}}\right)^2\right],
\end{align}
computable in terms of bilocal correlators that have been computed previously in \cite{Mertens:2019tcm}.\footnote{Upon Wick-rotating our $f \to if$ to go from the Lorentzian signature operator to the Euclidean one.} \\
There is a very natural extension to a non-abelian matter sector \cite{Mertens:2019tcm} that we postpone to future work.

\section{Fermi-Dirac spectrum for a Majorana fermion}
\label{app:fermion}

\subsection{Fermion field mode expansion in AdS$_2$}
In \cite{Anninos:2019oka}, a complex (Dirac) two-component spinor in the global AdS$_2$ frame was studied. We focus instead on a real (Majorana) spinor, and generalize to the arbitrary frame \eqref{bulkmetric}. We follow the conventions of \cite{Anninos:2019oka} and use the gamma matrices $\gamma^0 = i \sigma_1$ and $\gamma^1 = \sigma_3$, together with the massless Majorana equation:
\begin{equation}
\label{majeq}
i \gamma^a e^\mu_a \mathcal{D}_\mu \psi = 0, \qquad \qquad \mathcal{D}_\mu \equiv \partial_\mu +  \frac{1}{8}\omega^{ab}_\mu \left[\gamma_a, \gamma_b\right].
\end{equation}
Using lightcone coordinates $u=t+z$ and $v=t-z$ for the local Lorentz frame, and denoting the conformal factor of the metric as $\Omega^{2}\equiv \frac{(F(u)-F(v))^2}{F'(u)F'(v)}$, we can read off the zweibein and spin connection of the metric \eqref{bulkmetric} as:
\begin{align}
e^a_\mu \equiv e^u_u = e^v_v = \Omega^{-1}, \quad \omega^{ab}_\mu dx^\mu \equiv \omega^{uv} = \frac{2\partial_u \Omega}{\Omega} du - \frac{2\partial_v \Omega}{\Omega} dv
\end{align}
We hence find explicitly $\mathcal{D}_\mu = \left(\begin{array}{cc} \partial_\mu & -\frac{i}{4} \omega^{uv}_\mu \\ \frac{i}{4} \omega^{uv}_\mu & \partial_\mu \end{array}\right)$, and the equation decouples into two components for $\psi^\pm = \frac{1}{\sqrt{2}}(1+\gamma^0)\chi^\pm$ as
\begin{equation}
(\partial_u - \frac{1}{4}\omega^{uv}_u ) \chi^- = 0 , \quad (\partial_v + \frac{1}{4}\omega^{uv}_v ) \chi^+ = 0,
\end{equation}
solvable by the mode expansions:
\begin{equation}
\psi^+ = e^{-\frac{1}{4}\int^v dv \, \omega^{uv}_v }  \sum_\omega s_+ e^{- i \omega u} a_\omega^+, \qquad \psi^- = e^{\frac{1}{4}\int^u du \, \omega^{uv}_u } \sum_\omega s_- e^{- i \omega v} a_\omega^-,
\end{equation}
in terms of the spinors $s_+ = \frac{1}{\sqrt{2}}\left(\begin{array}{c} 1 \\ i \end{array}\right)$ and $s_- = \frac{1}{\sqrt{2}}\left(\begin{array}{c} i \\ 1 \end{array}\right)$. Combining both components, and explicitly plugging in the spin connection, we can write the full mode expansion as:
\begin{equation}
\label{modefermi}
\psi(u,v) \equiv \psi^+ + \psi^- = \left(\frac{F(u)-F(v)}{\sqrt{F'(u)F'(v)}}\right)^{1/2}\left[s_+ \sum_{\omega} e^{- i \omega u} a_\omega^+ + s_- \sum_{\omega} e^{- i \omega v} a_\omega^-\right],
\end{equation}
with $a_\omega^{\pm \dagger} = a_{-\omega}^{\pm}$ satisfying $\left\{a_\omega^\pm, a_{\omega'}^{\pm \dagger}\right\} = \delta(\omega-\omega')$. The components $\psi^\pm$ are the left- and right-handed Majorana-Weyl components. Up to the conformal prefactor $\Omega^{1/2}\equiv \left(\frac{F(u)-F(v)}{\sqrt{F'(u)F'(v)}}\right)^{1/2}$, these fields represent the flat space left- and right-moving degrees of freedom. \\
All of this can be viewed as a very explicit verification of the Weyl rescaling property of the massless fermion field equation \eqref{majeq} in 2d: given a solution $(g,\psi)$, the pair $(\Omega^{-2}g, \Omega^{1/2}\psi)$ is also a solution, and this is indeed the mode expansion \eqref{modefermi} constructed above.\footnote{
We also remark that Dirichlet boundary conditions at $z=0$ have to be imposed, basically setting $a_\omega^+ =- a_\omega^-$, and removing half of the oscillators. The Weyl rescaling mentioned here keeps fixed the holographic boundary, and preserves Dirichlet boundary conditions. Just as for the bosonic case, the computation in the next subsection can be done with any of the four options (taking $u$ or $v$ for each of the two operators in the two-point function), all of which would give the same outcome. For concreteness, we take $u$ in both cases.}

\subsection{Fermion number operator and occupation}
\begin{figure}[h]
\centering
\includegraphics[width=0.55\textwidth]{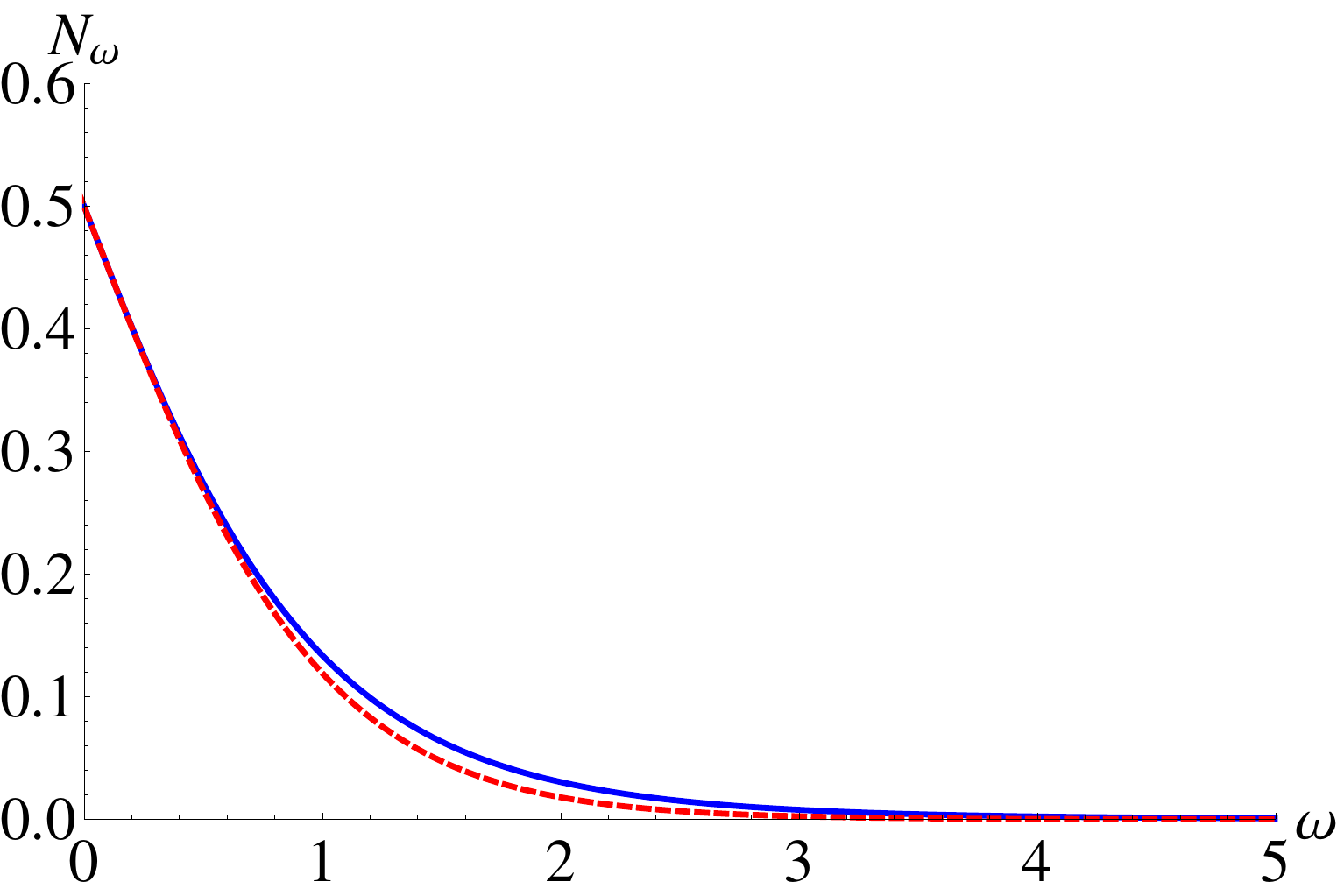}
\caption{Blue (upper): exact occupation number $\left\langle N_\omega \right\rangle_{\beta}$ of the Unruh radiation, computed from \eqref{exactfermi} with $\beta=2$. Red (lower): semi-classical fermionic (Fermi-Dirac) spectrum of Unruh radiation, coming from \eqref{semifermi}.}
\label{PlanckFermi}
\end{figure}

\begin{figure}[h]
\centering
\includegraphics[width=0.55\textwidth]{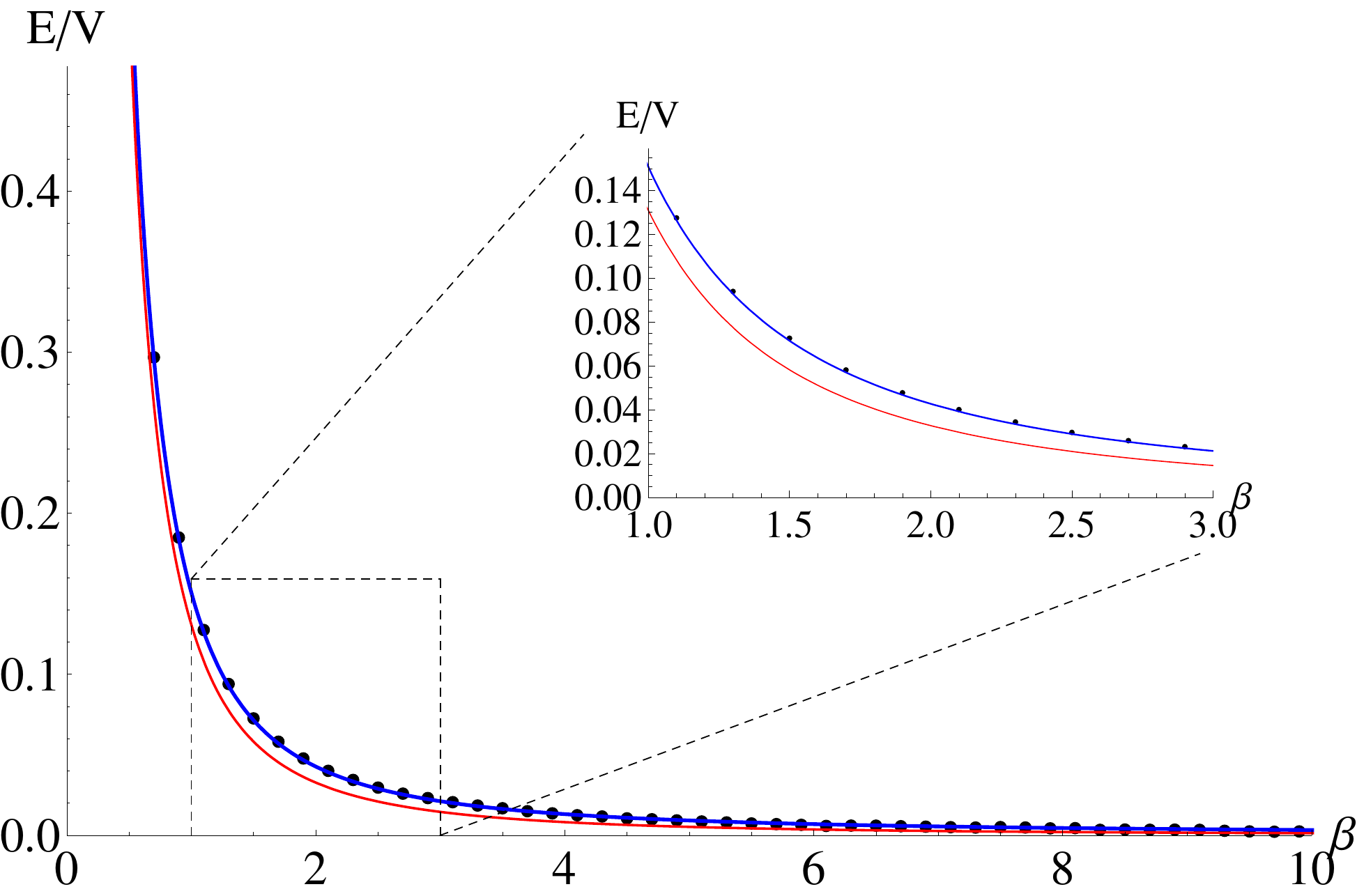}
\caption{Total energy density $\frac{1}{V}\int_{0}^{+\infty}d\omega \, \omega \left\langle N_\omega \right\rangle_{\beta}$ of the Unruh radiation, as a function of $\beta$, computed by integrating \eqref{exactfermi} (black dots). The exact energy \eqref{exUnruh} (with $c=1/2$) is plotted as a blue line (top). The semi-classical energy \eqref{Enmat} is plotted as a red line (bottom), computable by integrating \eqref{semifermi}. The inset shows in more detail the match at the exact level, and the approximation made by taking the semi-classical result.}
\label{PlanckFermiEbeta}
\end{figure}

Using the Weyl-transformation property of the fermion two-point function:
\begin{equation}
\label{weyltf}
\left\langle \psi^\pm_1 \psi^\pm_2\right\rangle_{\Omega^{-2} \eta} = \Omega^{1/2}_1 \Omega^{1/2}_2 \left\langle \psi^\pm_1 \psi^\pm_2\right\rangle_{\eta}
\end{equation}
to relate the AdS$_2$ metric \eqref{bulkmetric} to the flat metric,\footnote{The Weyl anomaly cancels out in numerator and denominator.} and using the flat result
\begin{equation}
\label{occfer}
\left\langle 0_F\right|\psi^+_1 \psi^+_2\left|0_F\right\rangle_{\eta} = \frac{\sqrt{F'(u_1)F'(u_2)}}{F(u_1)-F(u_2)},
\end{equation}
one can write the expression for the fermion number occupation number as:
\begin{align}
N_\omega[f] &= \left\langle 0_F\right| a_\omega^{+\dagger}a^{+}_\omega \left|0_F\right\rangle = \frac{1}{4\pi^2}\int du_1 \int du_2 \Omega_1^{-1/2}\Omega_2^{-1/2}e^{- i \omega (u_1-u_2)} \left\langle s_+^{\dagger}\psi^+_1 \psi_2^{+\dagger} s_+ \right\rangle_{\Omega^{-2} \eta} \nonumber \\
&= \frac{\omega}{\pi}\int dy_1 \int dy_2 u_\omega(y_1) u^*_\omega(y_2) \frac{\sqrt{f'(y_1)f'(y_2)}}{\frac{\beta}{\pi}\sinh \frac{\pi}{\beta}(f_1-f_2)},
\end{align}
where in the second line, we canceled the explicit conformal prefactors in these expressions, and wrote the expression in terms of the canonically normalized scalar modes $u_\omega(y) \equiv \frac{1}{\sqrt{4\pi \omega}}e^{-i\omega y}$.
\\~\\
Subtracting the vacuum contribution, we find that the number operator for a single Majorana fermion is given by:
\begin{align}
N_\omega[f] = \frac{\omega}{\pi}\int dy_1 &\int dy_2 u_\omega(y_1) u^*_\omega(y_2) \left[\frac{\sqrt{f'(y_1)f'(y_2)}}{\frac{\beta}{\pi}\sinh \frac{\pi}{\beta}(f_1-f_2)} - \frac{1}{y_{12}}\right],
\end{align}
computable from the $\ell=1/2$ bilocal Schwarzian operator.\footnote{Note that, combining \eqref{weyltf} and \eqref{occfer}, we can also study the bulk-to-bulk fermionic two-point function as the product of three Schwarzian bilocal operators along the lines of \cite{Blommaert:2019hjr}. We will not do that here.} 
\\~\\
As for the bosonic case in section \ref{sect:bbspec}, we average over time-ordered and anti-time-ordered two-point correlators to obtain the exact quantum gravitational occupation number:
\begin{equation}
\label{exactfermi}
\big\langle N_\omega \big\rangle_{\beta} = \frac{e^{-\beta\omega}-1}{2\pi^2} \frac{1}{Z}\int d\mu(k_2) \sinh(2\pi \sqrt{\omega+k^2})e^{-\beta k^2}\Gamma\left(\frac{1}{2} \pm i k \pm i \sqrt{\omega + k^2}\right) + \frac{1}{2}.
\end{equation}
Semi-classically, using the Fourier transform
\begin{align}
\int_{-\infty}^{+\infty} dt \frac{1}{\frac{\beta}{\pi}\sinh(\frac{\pi}{\beta}(t \mp i \epsilon))}e^{-i\omega t} &= \frac{1}{2\pi} e^{\mp\frac{\beta}{2} \omega}\Gamma\left(\frac{1}{2} + i \frac{\beta}{2\pi} \omega\right) \Gamma\left(\frac{1}{2} - i \frac{\beta}{2\pi} \omega\right) \nonumber \\
&= \frac{e^{\mp\frac{\beta}{2}\omega}}{e^{\frac{\beta}{2}\omega} + e^{-\frac{\beta}{2}\omega}},
\end{align}
and its $\beta \to \infty$ limit, one indeed finds the Fermi-Dirac population statistics of the thermal gas:
\begin{equation}
\label{semifermi}
N_\omega[f(y)=y] \,=\,  \frac{e^{-\frac{\beta}{2}\omega}}{e^{\frac{\beta}{2}\omega} + e^{-\frac{\beta}{2}\omega}}.
\end{equation}
Both of these formulas are compared in Figure \ref{PlanckFermi}.
As for the bosonic case, a slightly higher occupation number is observed. Notice that the quantum corrections are small at low energies. This can be explained due to a competition between quantum gravity and Pauli repulsion, the latter preventing any major modification in the population of these largely occupied energy levels. \\
We can check again that it is consistent with the total energy $E$ as computed by integrating the stress tensor \eqref{exUnruh} for $c=1/2$ matter (Figure \ref{PlanckFermiEbeta}).
\noindent Including charge can be readily done by combining the analysis of this section with that of the previous appendix.

\section{Matter entanglement entropy in curved spacetime}
\label{sect:enta}
For a curved 2d metric in conformal gauge $ds^2= - e^{\omega}dU dV$, the entanglement entropy for the interval $(U_1,V_1)$ to $(U_2,V_2)$ is written as
\begin{equation}
\label{Scur}
S = \frac{c}{12}(\omega_1 + \omega_2) + \frac{c}{12} \ln \frac{(U_1-U_2)^2}{\delta_{I1}\delta_{I2}} + \frac{c}{12} \ln \frac{(V_1-V_2)^2}{\delta_{I1}\delta_{I2}},
\end{equation}
with $\delta_I$ a UV cut-off measured by the \emph{inertial} observer at the endpoints. This formula was suggested in \cite{Fiola:1994ir} as the curved space generalization of the entanglement entropy formula, where refering to local inertial quantities is indeed the most natural thing to do. In our specific case, due to the presence of a prefered boundary coordinate, it is convenient to refer to that observer's time instead. Here we illustrate that the renormalized versions of these formulas agree. \\
Denoting $U_i = f(u_i)$ and $V_i = f(v_i)$, and specifying to an interval with the second endpoint $(U_2,V_2$) on the boundary of AdS$_2$, we write the entanglement entropy \eqref{Scur} for the Poincar\'e vacuum as:\footnote{Note that all dependence on the endpoints drops out of this formula \cite{Callebaut:2018xfu}.}
\begin{equation}
%S = \frac{c}{12}\ln \frac{1}{(f(u_1)-f(u_2))^2} + \frac{c}{12} \ln \frac{(f(u_1)-f(u_2))^2}{\delta_I^2} + (u \leftrightarrow v).
S = \frac{c}{12}\ln \frac{1}{(f(u_1)-f(v_1))^2} + \frac{c}{12} \ln \frac{(f(u_1)-f(v_1))^2}{\delta_I^2}.
\end{equation}
Subtracting the reference entropy in the $(u,v)$-frame:
\begin{equation}
%S_{\text{ref}} = \frac{c}{12}\ln \frac{f'(u_1) f'(u_2)}{(f(u_1)-f(u_2))^2} + \frac{c}{12} \ln \frac{(u_1-u_2)^2}{\delta_I^2} + (u \leftrightarrow v),
S_{\text{ref}} = \frac{c}{12}\ln \frac{f'(u_1) f'(v_1)}{(f(u_1)-f(v_1))^2} + \frac{c}{12} \ln \frac{(u_1-v_1)^2}{\delta_I^2},
\end{equation}
we can write the renormalized entropy as
\begin{equation}
%S_{\text{ren}} = S - S_{\text{ref}} = \frac{c}{12}\ln \frac{(f(u_1)-f(u_2))^2}{\delta^2 f'(u_1) f'(u_2)} - \frac{c}{12} \ln \frac{(u_1-u_2)^2}{\delta^2} + (u \leftrightarrow v),
S_{\text{ren}} = S - S_{\text{ref}} = \frac{c}{12}\ln \frac{(f(u_1)-f(v_1))^2}{\delta^2 f'(u_1) f'(v_1)} - \frac{c}{12} \ln \frac{(u_1-v_1)^2}{\delta^2},
\end{equation}
in agreement with the formulas in the main text in sections \ref{sect:matent} and \ref{sect:evap}, in principle valid for whatever cutoff we like, but we of course specify to the boundary observer's cutoff $\delta$. \\
Ultimately, this equality follows from the fact that in JT gravity, we only consider frames related by chiral mappings to the Poincar\'e frame. Otherwise, and in other models, one has to resort to \eqref{Scur} for the curved space entanglement formula.

\end{document}